\documentclass[useAMS,usenatbib,usegraphicx]{mn2e}

\usepackage{graphicx}
\usepackage{float}
\usepackage{color}
\usepackage{natbib}

\title[Spectral study of Puppis A and nearby sources]{Radio spectral
characteristics of the supernova remnant Puppis A and nearby sources}

\author[Reynoso and Walsh]{E. M. Reynoso$^{1}$\thanks{Email:
ereynoso@iafe.uba.ar}, and
        A. J. Walsh$^{2}$\thanks{Email: andrew.walsh@curtin.edu.au}\\
        $^{1}$Instituto de Astronom\'\i a y F\'\i sica del Espacio (IAFE), Av.
        Int. G\"uiraldes 2620, Pabell\'on IAFE, Ciudad Universitaria, \\Ciudad
        Aut\'onoma de Buenos Aires, Argentina\\ $^{2}$International Centre for
        Radio Astronomy Research, Curtin University, GPO Box U1987, Perth, WA
        6845, Australia}

\begin{document}

   \date{Received 20 March 2015; accepted 18 May 2015}

\pagerange{\pageref{firstpage}--\pageref{lastpage}} \pubyear{2015}

\maketitle

\label{firstpage}

\begin{abstract}

This paper presents a new study of the spectral index distribution of the 
supernova remnant (SNR) Puppis A. The nature of field compact sources is
also investigated according to the measured spectral indices. This work is 
based on new observations of Puppis A and its surroundings performed with the 
Australia Telescope Compact Array in two configurations using the Compact Array 
Broad-band Backend centred at 1.75 GHz. We find that the global spectral index 
of Puppis A is $\alpha = -0.563 \pm 0.013$. Local variations have been detected,
however this global index represents well the bulk of the SNR. At the SE, we 
found a pattern of parallel strips with a flat spectrum compatible with 
small-scale filaments, although not correlated in detail. The easternmost 
filament agrees with the idea that the SNR shock front is interacting with an 
external cloud. There is no evidence of the previously suggested correlation 
between emissivity and spectral index. A number of compact features are proposed
to be evolved clumps of ejecta based on their spectral indices, although 
dynamic measurements are needed to confirm this hypothesis. We estimate precise 
spectral indices for the five previously known field sources, two of which are 
found to be double (one of them, probably triple), and catalogue 40 new sources.
In the light of these new determinations, the extragalactic nature previously 
accepted for some compact sources is now in doubt.

\end{abstract}


\section{Introduction}\label{Int}

Puppis A is one of the most interesting supernova remnants (SNRs) in the 
Southern hemisphere. Its large extent ($\sim 50$ arcmin in diameter) and 
high surface brightness provides an unparalleled chance to study detailed
structures. Together with Cassiopeia A and G292.0+1.8, it is one of the three 
oxygen-rich Galactic SNRs, where fast oxygen knots of still uncontaminated
ejecta have been detected at optical wavelengths \citep{WK1985} and in X-rays
\citep{katsuda+08}.  Relative abundances of metal-rich ejecta measured on the 
NE portion are consistent with a progenitor with a mass in the range from 15 
to 25 M$\odot$ \citep{hwang+08, katsuda+10}. 

The compact X-ray source RX J0822-4300, near the centre of the shell and hence 
dubbed `central compact object' (CCO), has been identified as the stellar 
remnant left behind after the supernova (SN) outburst \citep{PBW96, ZTP99}. The 
explosion site was spotted by \citet{Wink+1988} through measuring proper 
motions of ejecta knots and filaments. This position is fully consistent with 
the direction of the CCO velocity \citep{HB2006,wp2007,Becker+2012}. 
\citet{gabi+06} propose that the CCO could be ejecting two opposite jets whose 
termination shocks appear in radio and X-rays as the eastern and western `ears'.

The age of Puppis A is estimated to vary between 3700 yr, based on the proper 
motion of optical filaments \citep{Wink+1988}, and 5200 yr, based on the 
velocity of the CCO and its projected distance to the explosion centre 
\citep{Becker+2012}. The brightest, eastern rim of the shell is apparently
interacting with a molecular cloud \citep{gd+ma88,EMR+95,paron+08}, whose
systemic velocity of $\sim +14$ km s$^{-1}$ translates into a kinematic 
distance of 2.2 kpc. However, \citet{beate+00} observed several pointings 
towards Puppis A and in the immediate vicinity in the four 18 cm lines of OH 
and concluded that the systemic velocity of the remnant is 7.6 km s$^{-1}$ 
rather than +14 km s$^{-1}$, which would bring the SNR about 1 kpc closer. 

In X-rays, Puppis A is one of the brightest SNRs, and the emission is mostly
dominated by shocked interstellar medium \citep[ISM;][]{BEK+05,hwang+08,Xgd+13}.
Infrared (IR) observations carried out with the {\it Spitzer Space Telescope}  
\citep{Spitzer+10} show an extremely good correlation with X-ray emission,
indicating that the thermal IR emission mostly arises from swept-up dust  
collisionally heated by the hot, shocked plasma. Puppis A has also been 
detected in $\gamma$-rays between 0.2 and 100 GeV with the {\em Fermi} satellite
\citep{Fermi+12}, but not above 260 GeV \citep{HESS-Pup}. The lack of emission 
at TeV energies implies that a spectral break or cutoff would occur at 280 or 
450 GeV, depending upon the power law model assumed. A break or radiative 
cutoff are expected if the acceleration of particles after the collision with 
an external cloud has now ceased \citep{HESS-Pup}.

Puppis A was widely observed at radio-continuum wavelengths, both with single 
dish telescopes \citep{MH69,kundu,milne71,MSH93} and interferometers 
\citep{AJG71,MGD83, Arendt+90,gd+91,gabi+06}. Analysis of the polarized 
emission \citep{kundu,milne72, dm76,MSH93} shows a predominantly radial 
magnetic field, low polarization, and little depolarization over most of the 
remnant. The first spectral index estimation of Puppis A was done by 
\citet{MH69} who, based on the computed value of $-0.5$, concluded that this 
radio source was an SNR. Subsequent studies refined the determination of the 
bulk spectral index as $\alpha = -0.52\pm 0.03$ \citep{gabi+06}\footnote{Note 
that, as an intermediate step in the data processing, \citet{gabi+06} used two 
flux densities, at 86 and 408 MHz, and found that $\alpha = -0.6$.} or $\alpha 
\simeq -0.53$ \citep{milne71,gd+91}. High-resolution studies have unveiled that
the spectral index varies spatially across the remnant \citep{gd+91,gabi+06}. 
In this paper, we revisit the spectral distribution of Puppis A applying a T-T 
plot method, which removes the problem of the zero level that can distort the
results if the background emission is not properly accounted for. Our
analysis is based on new wide band radio observations performed with the
Compact Array Broad-band Backend (CABB) of the Australia Telescope Compact 
Array (ATCA). In addition, we re-analyse five previously known compact sources
and list about 40 new compact sources in the field.

\section{Observations and data reduction}\label{Obs}

\begin{figure*}
\centering
\includegraphics{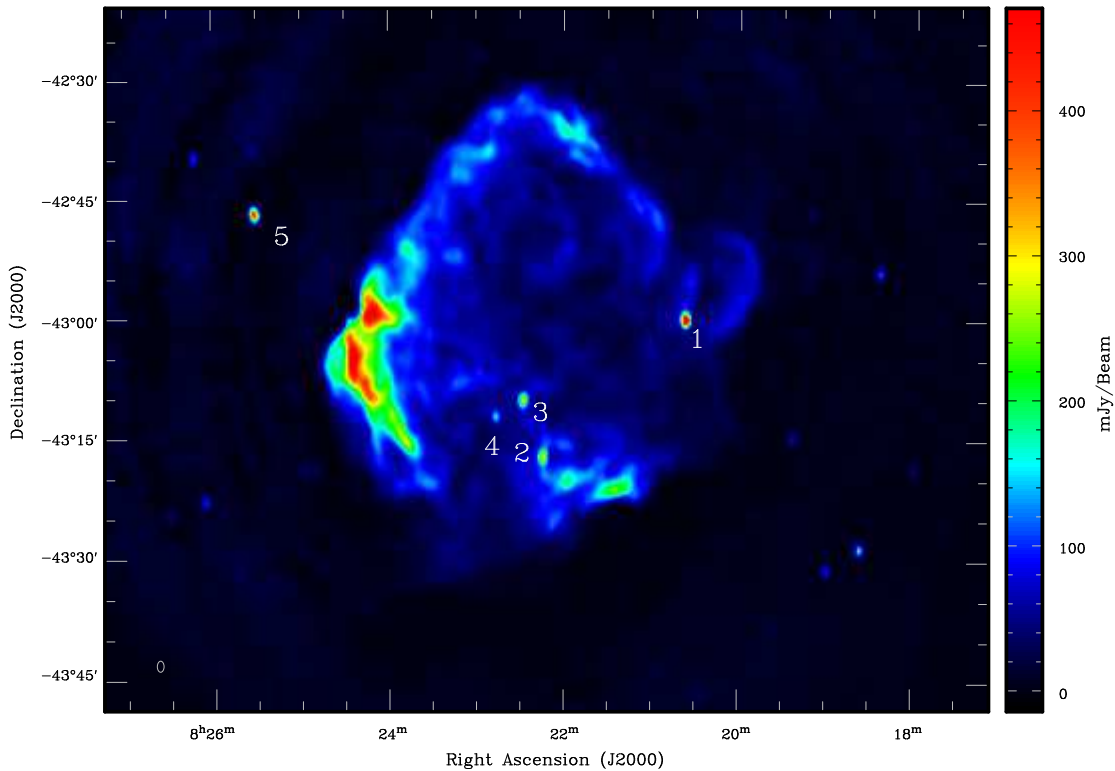}		
   \caption{Image of Puppis A at 1.4 GHz. The beam size is 82.2 $\times$ 50.6 
arcsec, with a position angle of -0\hbox{$.\!\!{}^\circ$}55 degrees. The flux 
density scale is shown at the right. The noise level is 1.5 mJy beam$^{-1}$. 
The beam is plotted at the bottom left corner. The white numbers 1 to 5 
correspond to the compact sources discussed in \S \ref{spind-cs}.
		}
     \label{ContMos}
\end{figure*}
Puppis A was observed simultaneously in the H{\sc i} 21 cm line, in four OH 18 
cm lines, and in radio-continuum with the CABB of the ATCA in two 13 hr runs. 
The first run took place on 2012 May 20 using the array configuration EW 352 
(baselines varying from 31 to 352 m excluding the 6th antenna), and the second, 
on 2012 July 6 in the 750A configuration (baselines from 77 to 735 m). Including
the 6th antenna adds baselines between 3015 and 3750 m for the 750A 
configuration, and between 4087 and 4439 for the EW 352 configuration. The 
primary flux and bandpass calibrator was PKS 1934-638, while PKS 0823-500 was 
used for phase calibration. Due to the large angular size of Puppis A, 
observations were made in mosaicking mode with 24 pointings in order to cover 
not only the SNR but also its surroundings. Following the Nyquist theorem, the 
pointings were separated by 19.6 arcmin to optimize the sampling at 1.4 GHz. 
The radio continuum was observed using a 2 GHz bandwidth divided into 2048 
channels of 1 MHz width, centred at 1750 MHz. The analysis of the H{\sc i} and 
OH lines will be developed in subsequent papers.

The data reduction was carried out with the {\sc miriad} software package 
\citep{Sault+95}. The {\em uv} data were split in bands of 128 MHz. For each 
band, calibrator and source observations were inspected and flagged for 
interference and corrupted data. The standard data calibration routines were 
applied. We have not used the whole set of visibilities to construct a map 
because the presence of sources with very different spectral slopes within the 
observed field will introduce non-negligible errors due to bandwidth smearing 
in a frequency band as wide as 2 GHz. Therefore, we preferred to use at most 
three consecutive bands in a single image. Using a multifrequency technique, we 
constructed an image at 1.4 GHz merging the bands centred at 1302, 1430 and 
1558 MHz. The data were Fourier-transformed with uniform weighting to 
reduce sidelobes, and cleaned with MOSMEM. The resulting image, shown in Fig. 
\ref{ContMos}, has a beam of 82.2 $\times$ 50.6 arcsec, with a position angle 
of $-0\hbox{$.\!\!{}^\circ$}55$, and a noise level of 1.5 mJy beam$^{-1}$, and 
is centred at 1395.6 MHz. A few residual sidelobe rings from the bright compact 
source at RA(2000)=8$^{\mathrm h}$20$^{\mathrm m}$35\hbox{$.\!\!{}^{\rm s}$}6,
Dec.(2000)=$-43^\circ$0\hbox{$^{\prime}$}23\hbox{$^{\prime\prime}$} could not 
be completely removed. The integrated flux density is computed to be 115 $\pm$ 8
Jy, where the error is dominated by the uncertainty in the background emission 
and the outermost contour level adopted. The agreement between this value and 
that reported by \citet{gabi+06} indicates that at this frequency the 
{\em uv}-coverage, combined with the mosaicking technique, samples all scale 
structures within Puppis A. 

To compute spectral indices, we applied a standard T-T plot method (see \S 
\ref{spind}). This method requires that the two images to be compared, one at 
each frequency, be obtained from visibilities within a same {\em uv} range and 
convolved with the same beam. The first condition ensures that both images 
sample features with similar scales. Adopting the first frequency to be 1.4 GHz 
and the second 2.5 GHz, visibilities were found to overlap in the range between 
0.2 and 4 k$\lambda$. We therefore constructed an additional image at 1.4 GHz 
following the same method as in Fig. \ref{ContMos} but using only this 
limited visibility range. The {\em uv}-filtered image thus obtained has a beam 
of 80.8 $\times$ 49.5 arcsec, with a position angle of 
$-0\hbox{$.\!\!{}^\circ$}4$. The image at 2.5 GHz was constructed merging the 
visibilities within the same {\em uv} range (0.2-4 k$\lambda$) in the bands 
centred at 2326, 2454 and 2582 MHz. This image was convolved with the same 
beam as the {\em uv}-filtered image at 1.4 GHz. To enhance the compact sources 
reducing the diffuse emission, two more sets of maps were formed with more 
restrictive {\em uv} ranges, from 0.5 to 4 k$\lambda$ and from 1 to 4 
k$\lambda$, respectively.

Finally, to perform an accurate analysis of the compact sources, an image was 
constructed for each band applying a more restrictive visibility range, from 0.8
to 4 k$\lambda$, to filter out extended emission. These images were then used 
to compute flux densities and sizes, minimizing the uncertainties introduced by
underlying background emission. The reduction in the number of visibilities due 
to the use of one single band and the restricted {\em uv} range results in a 
poor imaging and a higher noise, which makes the recognition of the weakest
sources more difficult. Such sources were confirmed when they were detected in 
most frequency bands.

\section{Results} 

\begin{figure}
\centering
\includegraphics[width=7.5cm]{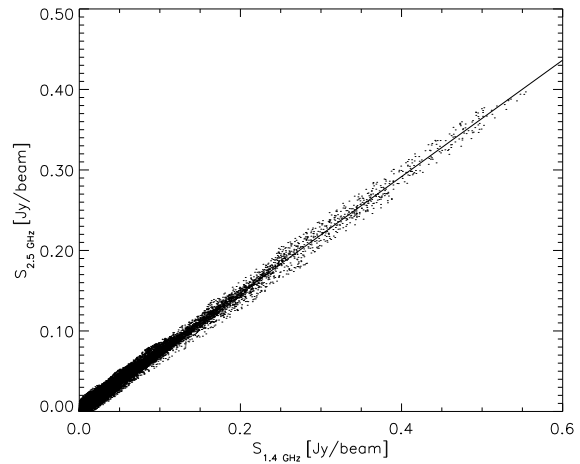}
   \caption{T-T plot between 1.4 and 2.5 GHz for the whole SNR Puppis A. Clip
levels of 15 and 10 mJy beam$^{-1}$, respectively, were applied. Background 
sources were removed. The fit yields a spectral index of $\alpha=-0.573 
\pm 0.004$.}
     \label{TT-Pup}
\end{figure}
\subsection{Spectral index computation}\label{spind}

\begin{table*}
\caption{Catalogued compact sources}
\label{cat-cs}
\centering
\begin{tabular}{cccrrr}
\hline
\hline
Source& RA (J2000) & Dec. (J2000) &~Spectral &~Spectral & Previous\\
number& ~~h~~ m ~~ s & ~~$^\circ$ ~~~$^\prime$ ~~~$^{\prime\prime}$ & 
index$^{\rm a}$ & index$^{\rm b}$ & names\\
\hline

1 &08 20 35.6 & -43 00 23& -0.76$\pm$0.04 &-0.86$\pm$0.04&G260.2--3.7$^{(1)}$ \\
2 &08 22 13.4 & -43 17 38& -0.91$\pm$0.07 &-1.09$\pm$0.05 \\
3 &08 22 27.0 & -43 10 29& -0.26$\pm$0.07 &-0.56$\pm$0.5 & ATPMN J082226.8--431026$^{(2)}$\\
 & & & & & 2XMM J082226.9--431026$^{(3)}$\\
4 &08 22 45.7 & -43 12 38& -0.75$\pm$0.15 &-0.94$\pm$0.08&G260.7--3.7$^{(1)}$ \\
5 &08 25 30.7 & -42 46 57& -0.92$\pm$0.04 &-1.09$\pm$0.09&Cul 0823-426$^{(4)}$\\
\hline
\end{tabular}
\begin{list}{}{}
\item{$^{\rm a}$} T-T plot. $^{\rm b}$ Slope of $S(\nu)$ versus $\nu$.
\item{{\it Notes.} (1)} \citet{DCC72}; (2) \citet{ATPMN}; (3) \citet{ssXMM}; (4) \citet{Slee77}
\end{list}
\end{table*}

To compute spectral indices minimizing the uncertainty introduced by the
always arbitrary selection of the zero level, we have applied a T-T plot 
technique. A description of this method, widely used over selected regions in
extended radio sources \citep[e.g.][]{Costain60,AR1993,uy+04,leahy06,rg07}, 
can be found in \citet{Turtle62} and \citet{Reiches88}. Basically, the idea is 
to assume that the observed continuum emission at a given frequency $T(\nu)$ 
can be written as $T(\nu) = T_{\rm s}(\nu) + T_0(\nu)$, where $T_{\rm s}(\nu)$ 
is the source emission and $T_0(\nu)$ is an isotropic background component, 
which includes an offset zero. Then, if $T_{\rm s}(\nu) \propto \nu^{\beta}$ 
with $\beta$ constant, the brightness temperature at two different frequencies 
are linearly related as 
\begin{equation}\hskip 2 cm
T(\nu_1) = T(\nu_2) \left( \frac{\nu_1}{\nu_2} \right) ^{\beta} + 
C,
\end{equation}
where $C = T_{0}(\nu_1) - T_{0}(\nu_2) \left( {\nu_1 / \nu_2} \right) 
^{\beta}$. Thus, plotting $T(\nu_1)$ in terms of $T(\nu_2)$ for all pixels
within a given region, the value of the brightness temperature spectral index 
$\beta$ is readily obtained from the slope of the linear regression 
independently of the zero level of each of the two maps.  

In what follows, we will refer to the flux density spectral index $\alpha$
defined by $S(\nu) \propto \nu^{\alpha}$, which is related to the brightness 
temperature spectral index as $\alpha = \beta +2$. As the region where this 
method is applied becomes smaller, there is a better chance that both 
hypotheses, i.e. background isotropy and constant $\alpha$, approach reality. 
A limitation of this technique, already pointed out by \citet{AR1993}, is that 
spectral indices can only be measured at scales compatible with the region 
analysed, while larger features are filtered out. 

\begin{figure*}
\centering
\includegraphics{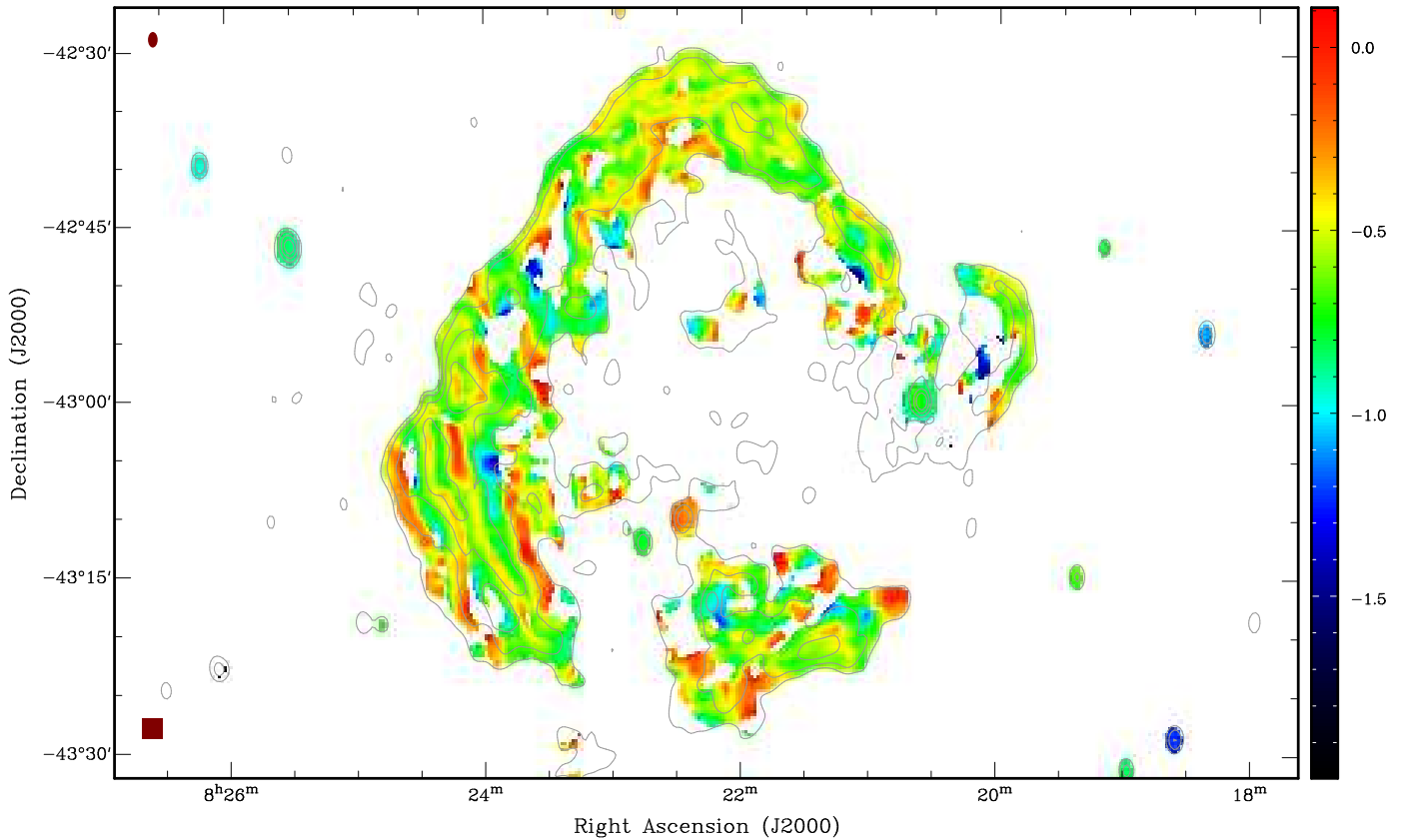}	
   \caption{Spectral index distribution along Puppis A and surroundings. The
{\bf filled} box of 7$\times$7 pixels at the bottom left corner represents the 
{\bf size of the} area where T-T plots were computed. All pixels where the 
linear fits to the T-T plot had a correlation coefficient lower than 0.950 were 
blanked. The bar at the right displays the spectral index. Radio continuum 1.4 
GHz contours at 20, 60, 150, 300 and 500 mJy beam$^{-1}$ are overlaid. The beam
is shown 
at the top left corner.}
     \label{alfa-map}
\end{figure*}

The quoted uncertainties correspond to the formal statistical errors of 
the linear regressions scaled by $\sqrt{\rm {(beam\  area/pixel\  area)}}$ 
\citep{green90}. The scaling factor is due to the oversampling of the images,
otherwise the statistical errors in the linear fits may be underestimated since 
adjacent pixels are not independent and there are typically many pixels 
per beam \citep[see for example ][]{gaens+99,dohert+03}. The quoted 
uncertainties, therefore, account for oversampling and contain possible 
small-scale variations of the local background emission. In this estimation, 
the contribution of the flux density noise levels of each frequency map has not 
been taken into account.

Although T-T plots are less accurate for extended sources, we used this method
to estimate a global spectral index for Puppis A. We first plotted the 
{\em uv}-filtered image at 2.5 GHz in terms of the same at 1.4 GHz (Fig. 
\ref{TT-Pup}). To avoid spurious contribution from emission unrelated to Puppis 
A, the brightest point sources projected within the SNR shell were removed in 
both images. The 1.4 and 2.5 GHz images were clipped at 15 and 10 mJy 
beam$^{-1}$, respectively. A linear fit yielded a spectral index of 
$\alpha=-0.573 \pm 0.004$, with a correlation coefficient of 0.9946. Inverting 
the frequencies, the result is $\alpha=-0.554 \pm 0.004$. Averaging both 
results, the global spectral index of Puppis A is $\alpha=-0.563 \pm 0.013$, 
where the error encloses the two values obtained above.

To map the spectral index distribution, we considered each pixel as the centre  
of a 7$\times$7 pixel box and fitted a T-T plot over the box between 1.4 and
2.5 GHz. The box was slid in two dimensions over the SNR at 1 pixel 
increments. We adopted clip levels of 20 and 15 mJy beam$^{-1}$ respectively, 
and all linear fits where the correlation coefficients were lower than 
0.950, or where the regression involved less than 6 pixels, were blanked out. 
This procedure was repeated for $S_{\rm 2.5 GHz}$ against $S_{\rm 1.4 GHz}$
and vice versa, and the average between both maps was computed.
The result is shown in Fig. \ref{alfa-map}. In good agreement with the global
spectral index estimated through the T-T plot shown in Fig. \ref{TT-Pup}, the
average spectral index of Puppis A obtained over Fig. \ref{alfa-map} after
removal of the background sources gives $\alpha = -0.59 \pm 0.22$, where the 
larger error is due to spatial variations of the synchrotron spectra.

\subsection{Field compact sources}\label{spind-cs}

A number of compact sources, either projected within the SNR shell or in the
surroundings, are present in our maps. However, only five of them have been
reported in previous surveys \citep{DCC72,Slee77, MGD83,gd+91}, four of them
superimposed on the SNR face. In Fig. \ref{ContMos}, these sources have been 
labelled with numbers increasing with Right Ascension \citep[note that][name 
them with letters A to E following the same ordering]{MGD83}.  According to 
\citet{MGD83} and \citet{gd+91}, sources 1 to 4 have spectral indices close to 
$-1.0$, pointing to an extragalactic origin. Those indices are based on the 
direct comparison of integrated flux densities at two frequencies. Source 5, 
catalogued as 0823-426 by \citet{Slee77} and misidentified by \citet{gd+91} as 
their source 4, did not have enough information to accurately compute a 
spectral index \citep{MGD83}. Moreover, it was observed at 80 and 160 MHz with 
the Culgoora radioheliograph \citep{Slee77} but was not identified in the 
Parkes survey at 2.7 GHz \citep{DCC72}. 

\begin{table}
\caption{Spectral variations for sources 3 and 4}
\label{curv-fits}
\centering
\begin{tabular}{cccc}
\hline
\hline
& Flux density range & Flux density range &~Spectral \\
& at 1.4 GHz (mJy) & at 2.5 GHz (mJy) & 
~index\\
\hline

 & $S \geq 45$ & $S \geq 35$ &-0.26$\pm$0.06 \\
3 & $45 \leq S \leq 100$ & $35 \leq S \leq 80$ &-0.5$\pm$0.3 \\
 & $S \geq 150$ & $S \geq 120$ &-0.14$\pm$0.20 \\
\hline
 & $S \geq 30$ & $S \geq 20$ &-0.7$\pm$0.2 \\
4 & $30 \leq S \leq 80$ & $20 \leq S \leq 50$ &-1.0$\pm$0.4 \\
 & $S \geq 80$ & $S \geq 50$ &-0.55$\pm$0.35 \\
\hline
\end{tabular}
\end{table}

\begin{figure}
\centering
\includegraphics[width=.23\textwidth]{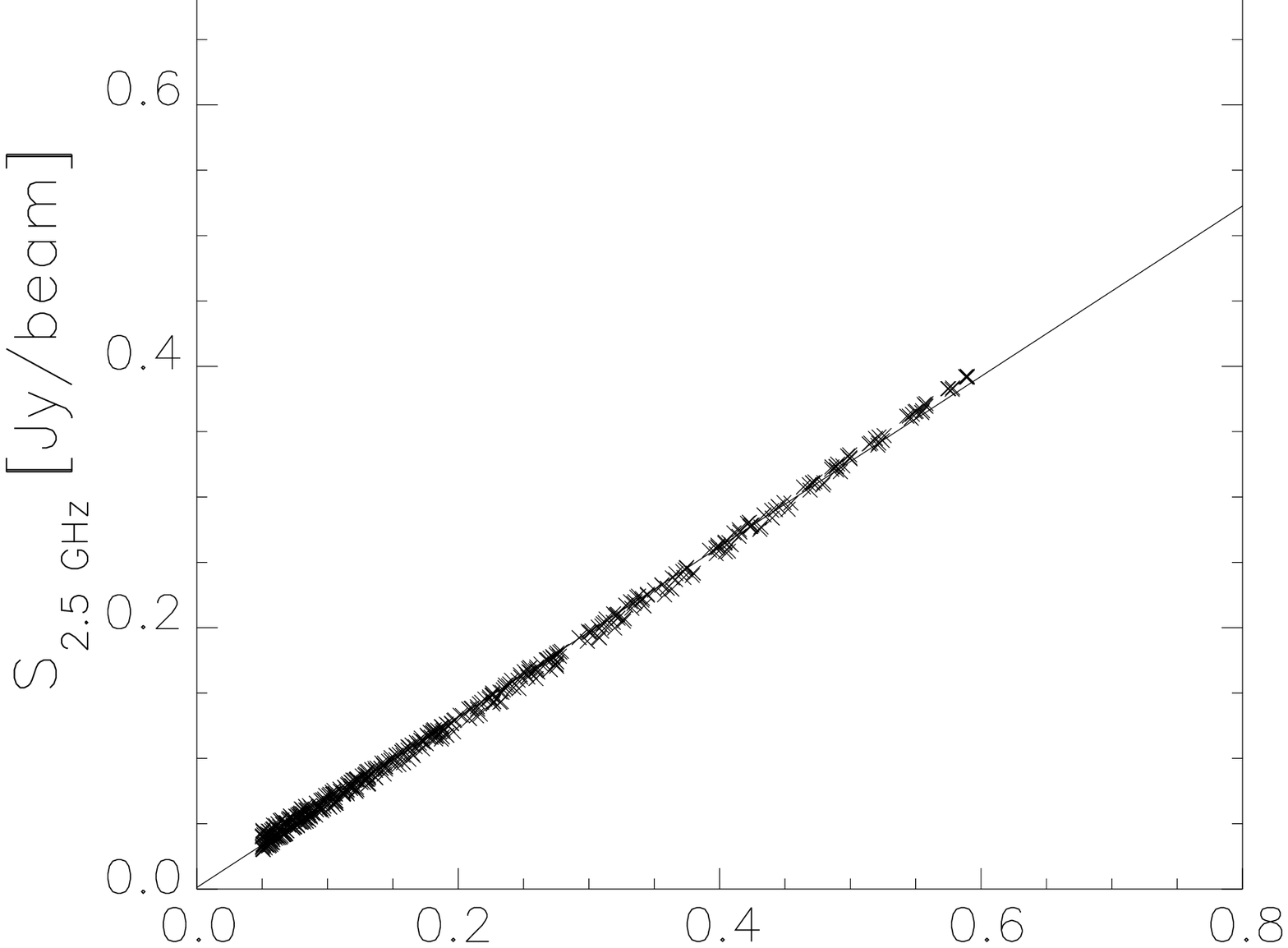}~\hfill
\includegraphics[width=.235\textwidth]{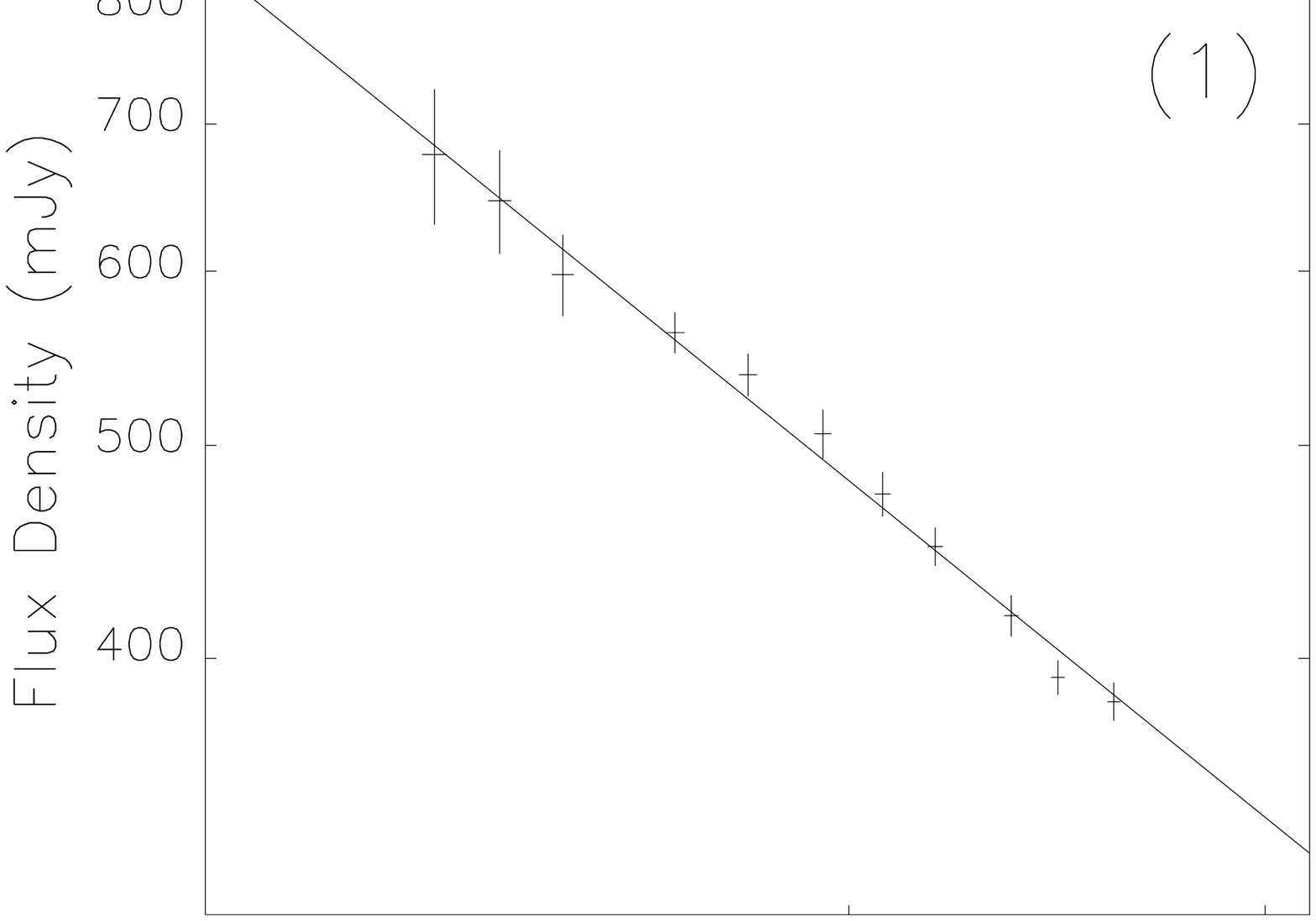}\\
\includegraphics[width=.23\textwidth]{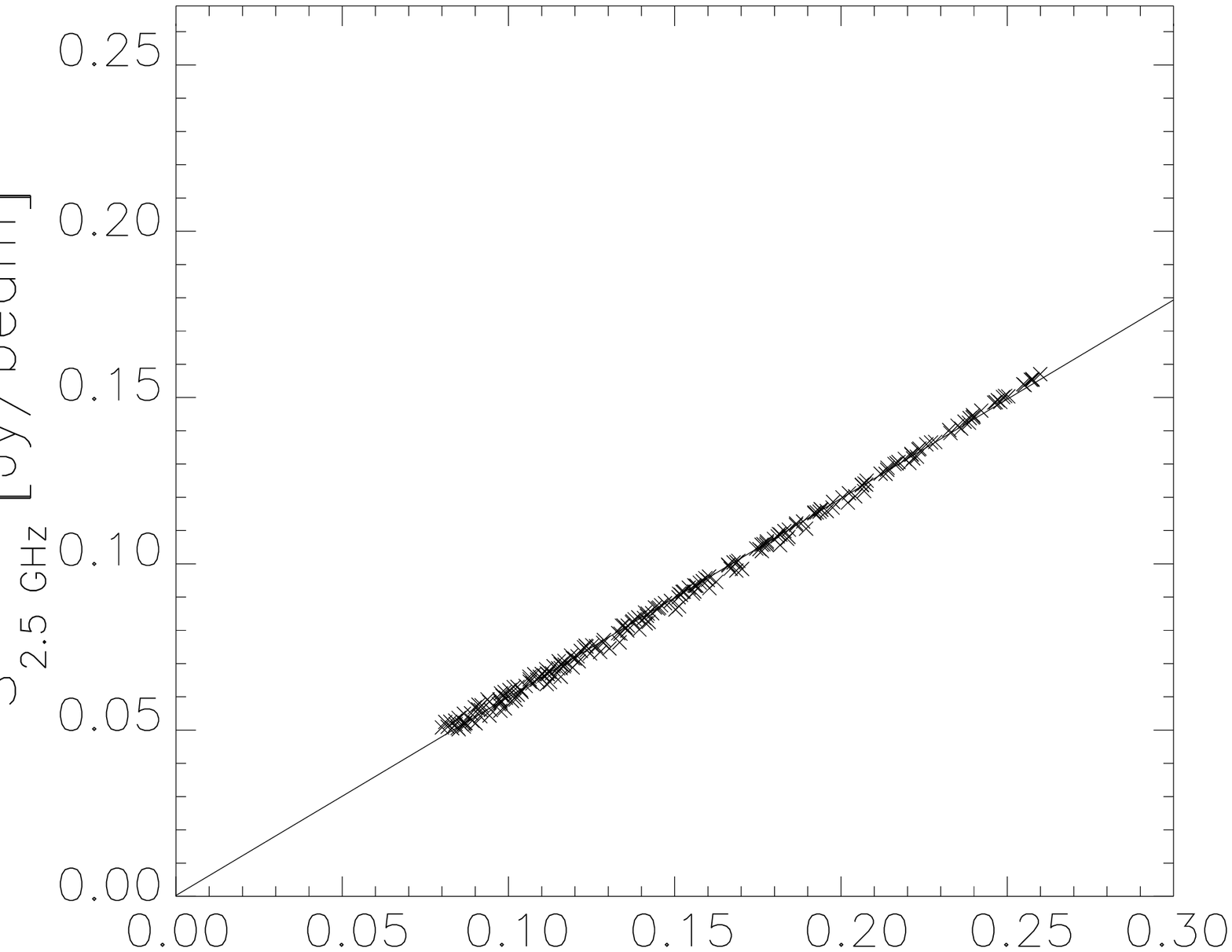}~\hfill
\includegraphics[width=.235\textwidth]{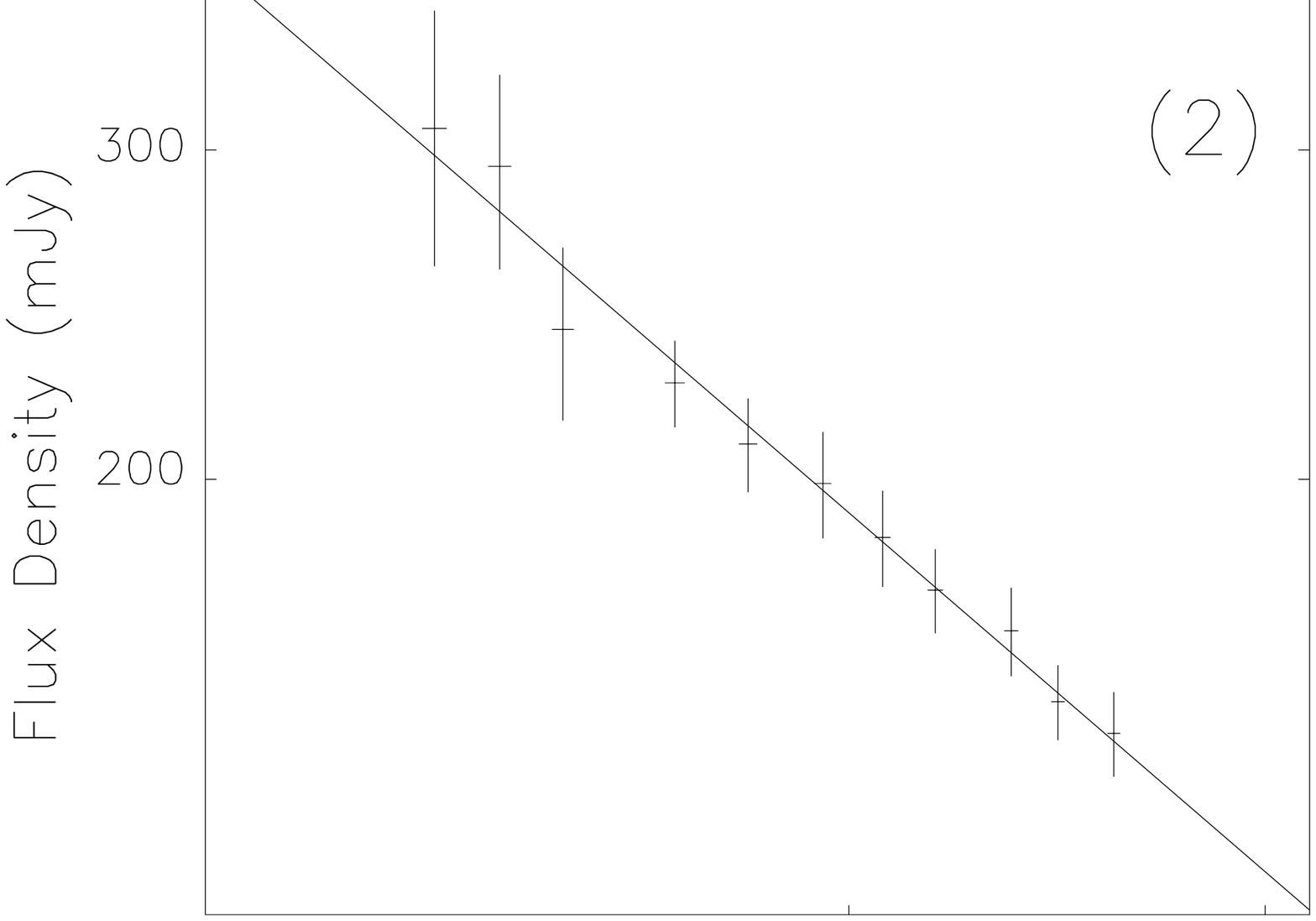}\\
\includegraphics[width=.23\textwidth]{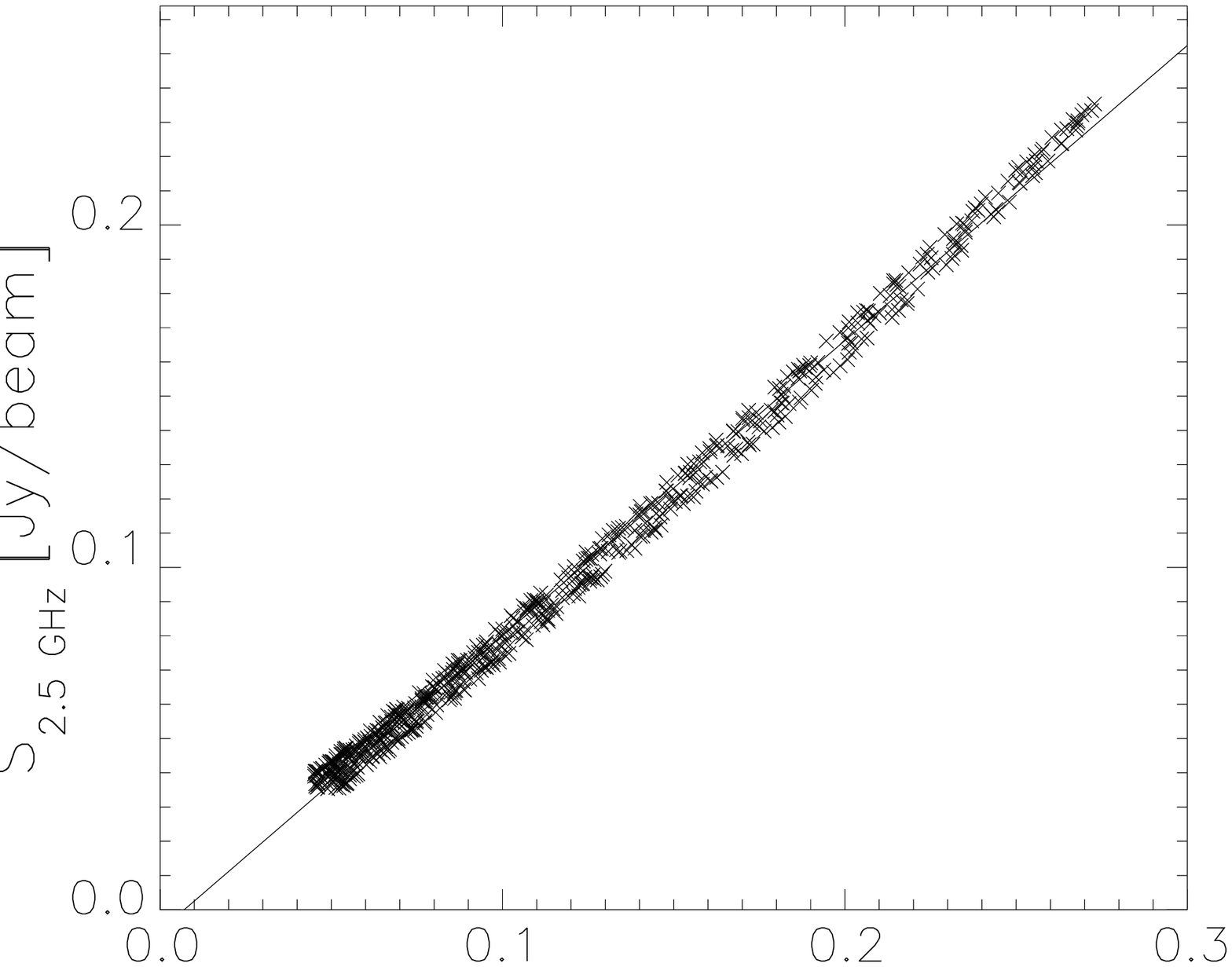}~\hfill
\includegraphics[width=.235\textwidth]{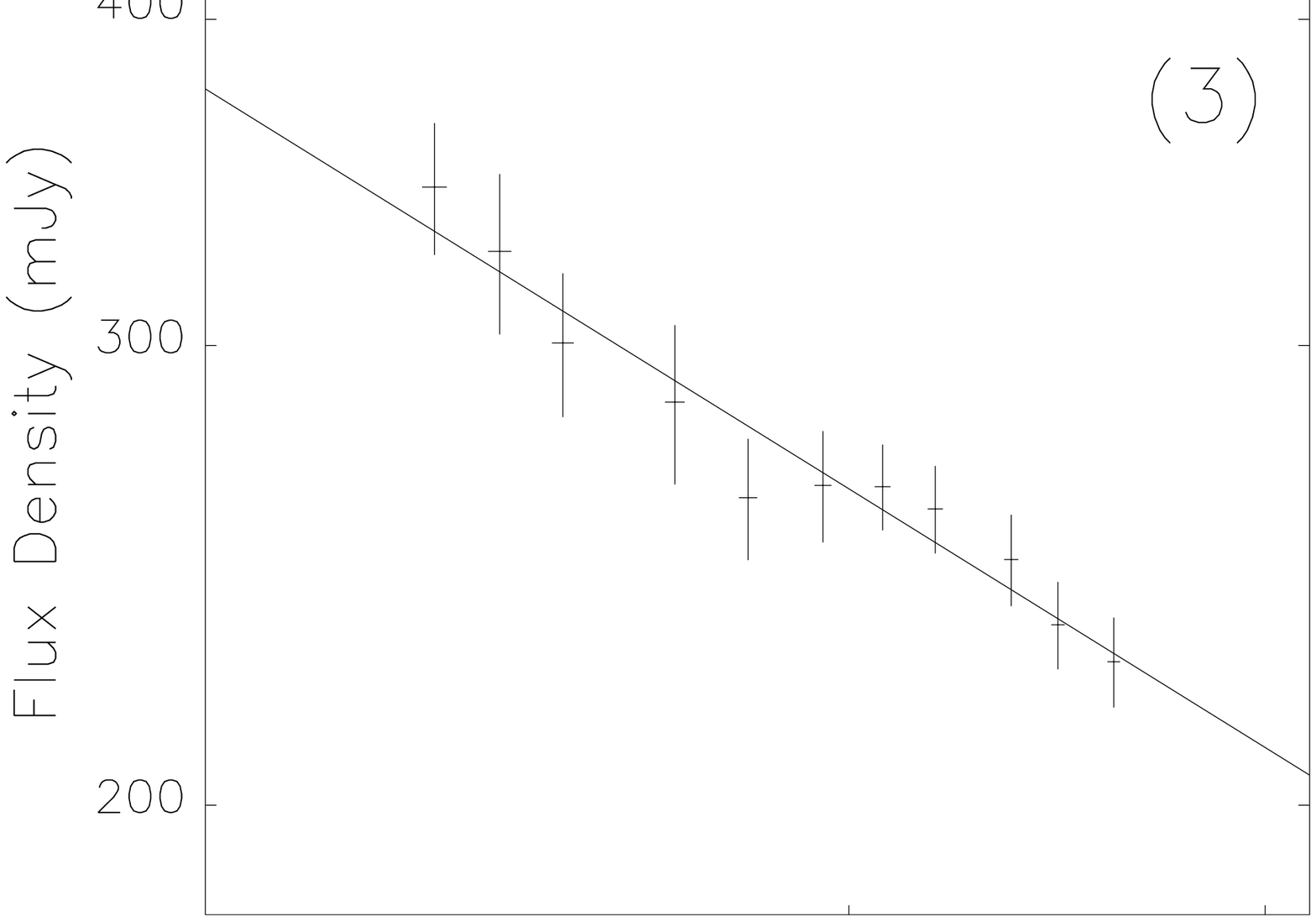}\\
\includegraphics[width=.23\textwidth]{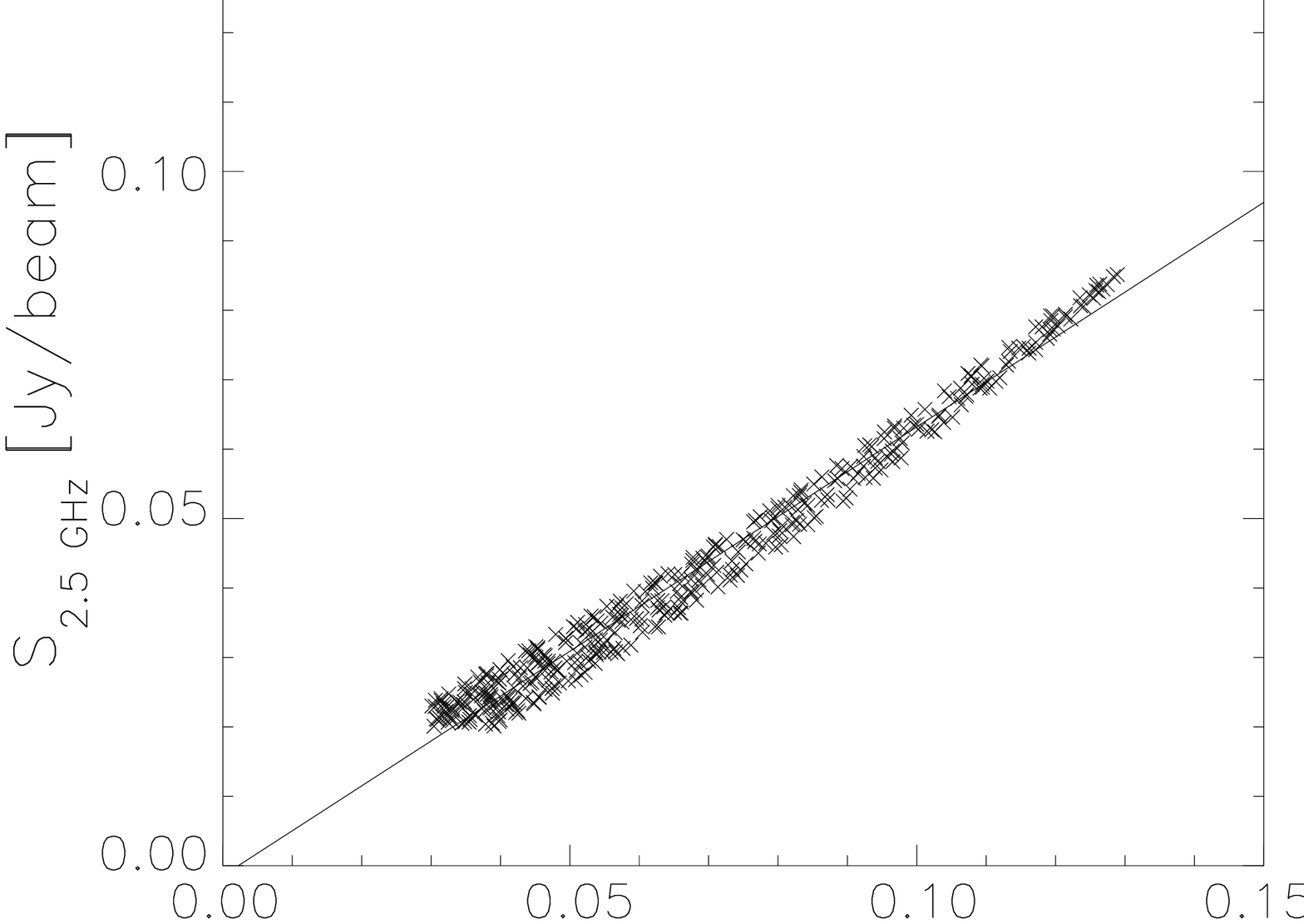}~\hfill
\includegraphics[width=.235\textwidth]{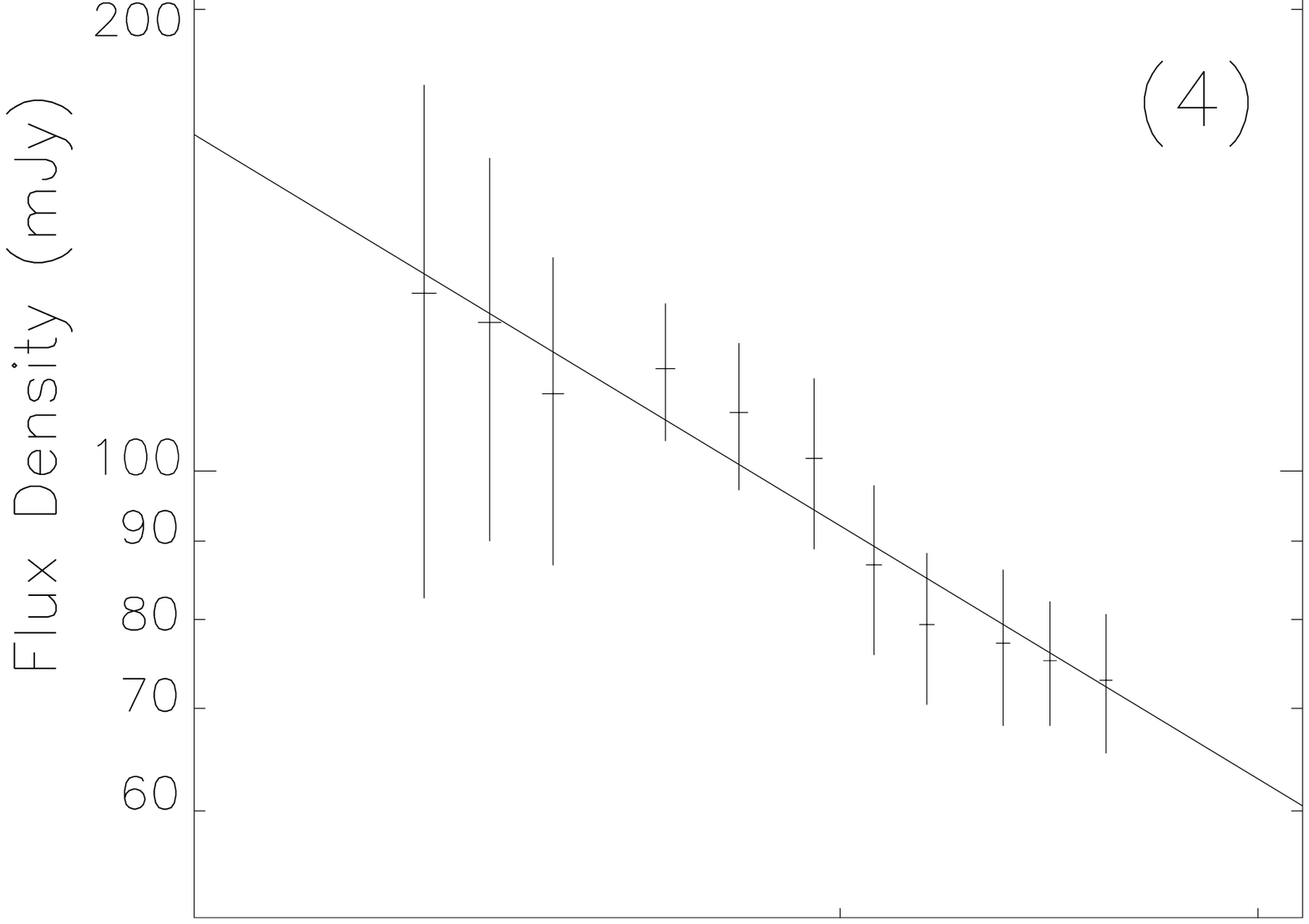}\\
\includegraphics[width=.23\textwidth]{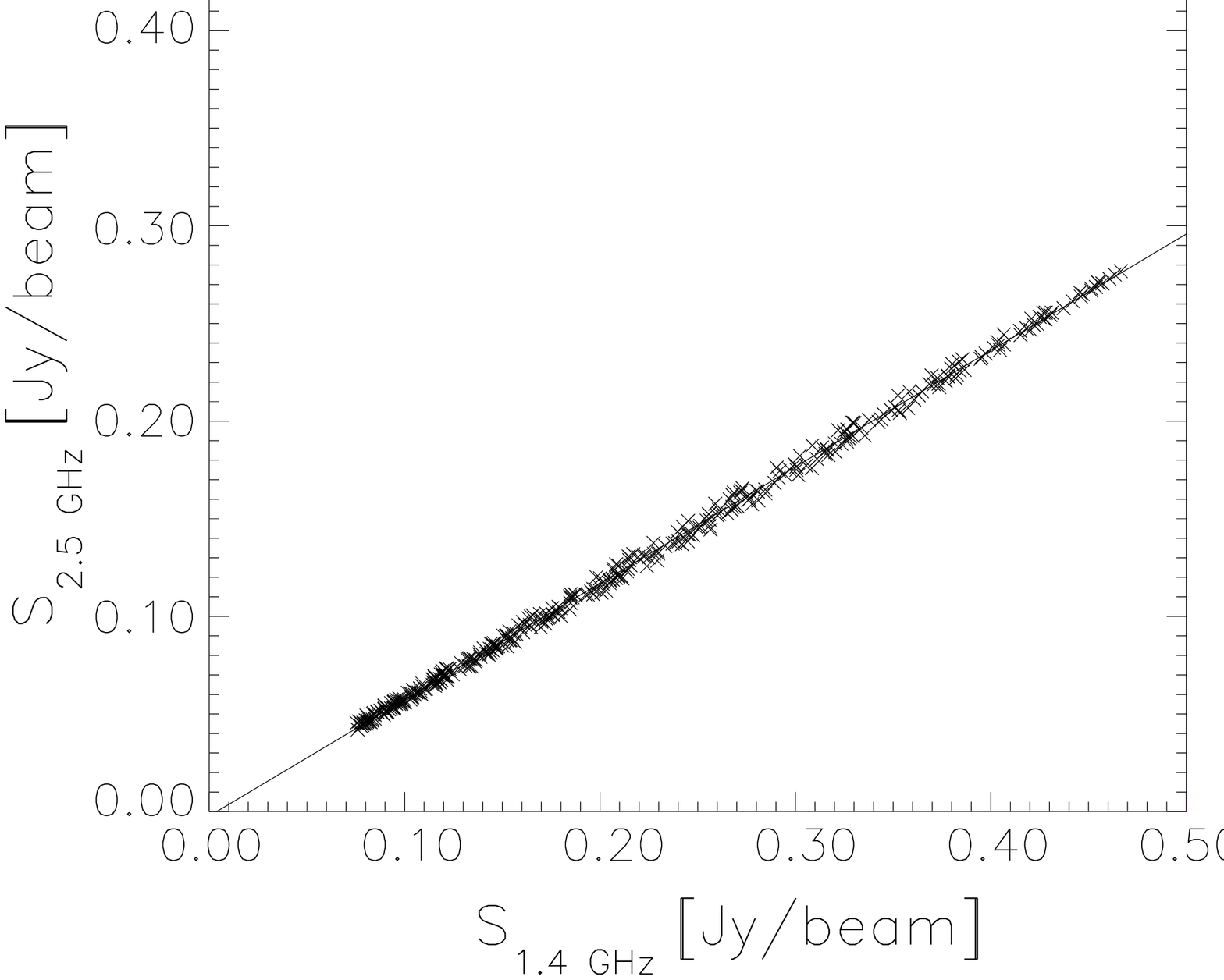}~\hfill
\includegraphics[width=.235\textwidth]{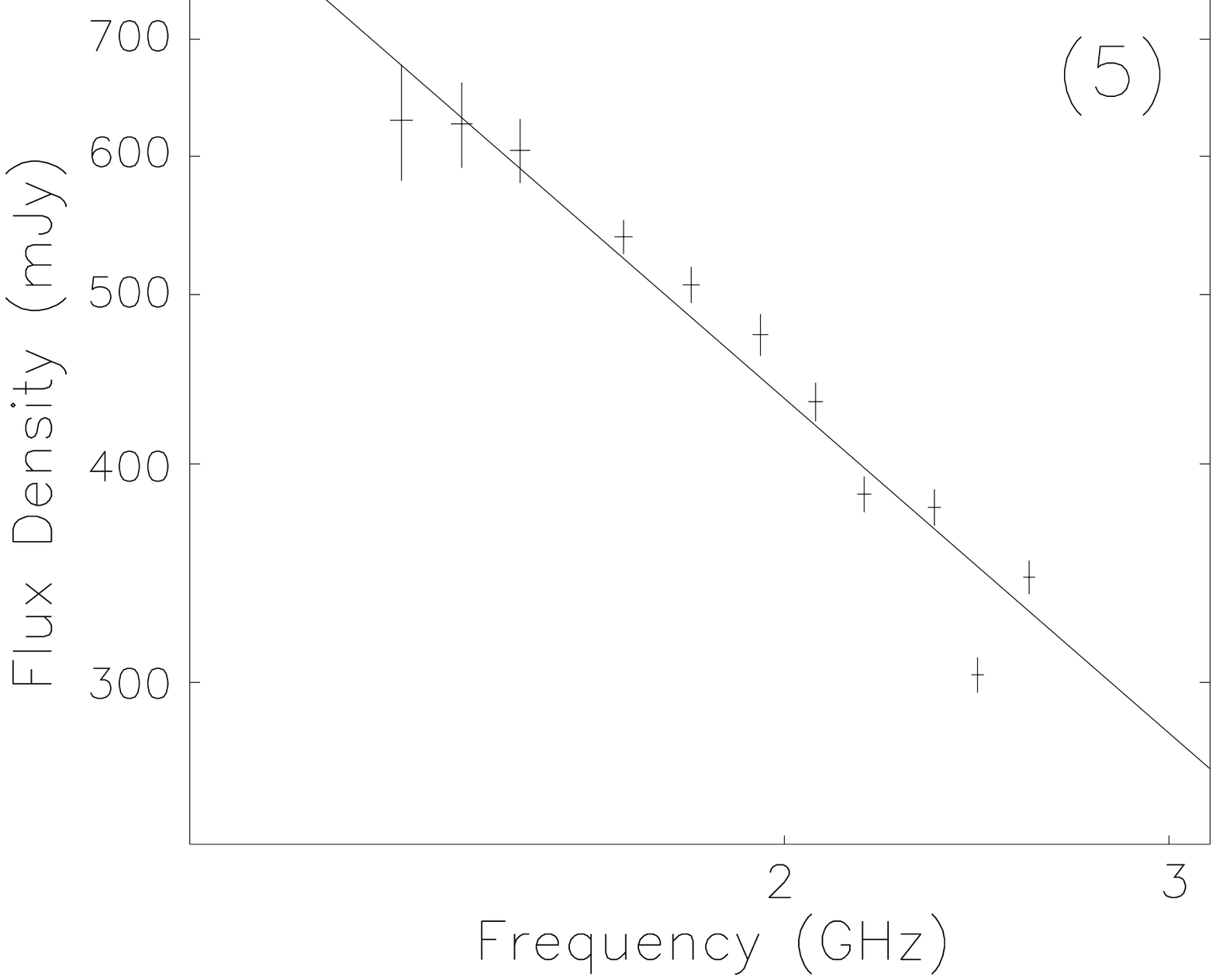}
\caption For each of the five previously known sources, T-T plots (left-hand
column) and the corresponding linear fits to the flux densities $S(\nu)$ in 
terms of frequency (right-hand column). The numbers at the top right corner 
of each frame in the right-hand column correspond to the number assigned to 
each source in Table \ref{cat-cs} and in Fig. \ref{ContMos}.
\label{plots-cs}
\end{figure}

\begin{table}
\caption{New compact sources}
\label{new-cs}
\centering
\begin{tabular}{cccrr}
\hline
\hline
& RA (J2000) & Dec. (J2000) & $S_{\rm 1.3 GHz} ^a$ & ~Spectral\\
& ~~h~~ m ~~ s & ~~$^\circ$ ~~~$^\prime$ ~~~$^{\prime\prime}$ & mJy & index \\
\hline

6 & 08 17  58.6 &  -43  18  37  &  53.5 $\pm$ 8.5 \\
7 & 08 18  22.6 &  -42  54  12.5  &  133 $\pm$ 7 & -1.25 $\pm$ 0.07 \\
8 & 08 18  35.3 &  -43  28  47.0  &  180 $\pm$ 10 & -1.17 $\pm$ 0.08 \\
9 & 08 18  42  &  -42  53  03  &   20 $\pm$ 6 & -0.5 $\pm$ 0.4\\
10& 08 18  57.9 &  -43  31  29.5  &  90 $\pm$ 7 & -0.78 $\pm$ 0.08  \\
11& 08 19  11 &  -42  46  55  &  46 $\pm$ 6 & -0.75 $\pm$ 0.25\\
12& 08 19  18.5  &  -43  23  55  &   10 $\pm$ 3.5 \\
13& 08 19  22.2 &  -43  15  09  &  76.5 $\pm$ 7.5 & -0.70 $\pm$ 0.15\\
14& 08 19  35  &  -43  39  03    &  17.5 $\pm$ 7.5 \\
15& 08 19  49  &  -43  24  40    &  14 $\pm$ 4 & -0.1 $\pm$ 0.3\\
16& 08 19  58  &  -43  50  50.3  &  49 $\pm$ 6\\
17& 08 20  25  &  -43  47  11    &  12 $\pm$ 6\\
18& 08 20  40  &  -42  09  14    &  29 $\pm$ 4\\
19& 08 20  51  &  -42  24  00  &  18 $\pm$ 6 & -0.74 $\pm$ 0.30\\
20& 08 21  09  &  -43  22  35.4  &  46 $\pm$ 3\\
21& 08 21  22  &  -42  16  19    &  40 $\pm$ 6 & -1.12 $\pm$ 0.25\\
22& 08 21  42  &  -42  31  43    &  33 $\pm$ 8 & -1.0 $\pm$ 0.4\\
23& 08 21  51  &  -43  48  37  &  35 $\pm$ 6\\
24& 08 22  20  &  -43  51  24  &  18 $\pm$ 6\\
25& 08 22  38  &  -42  30  19    &  29 $\pm$ 4 & -0.70 $\pm$ 0.45\\
26& 08 22  56 &  -42  26  55    &  36 $\pm$ 6 & -0.35 $\pm$ 0.45\\
27& 08 24  27 &  -43  00  24  &  85 $\pm$ 15 \\
28& 08 24  48  &  -43  19  24    &  26 $\pm$ 5 & -0.53 $\pm$ 0.45\\ 
29& 08 24  57  &  -43  19  17.3  &  37 $\pm$ 5 & -0.6 $\pm$ 0.3\\
30& 08 25  05  &  -43  09  21    &  12 $\pm$ 8\\
31& 08 25  06  &  -42  42  19    &  12 $\pm$ 5\\
32& 08 25  31  &  -42  39  09    & 27 $\pm$ 6 & -0.8 $\pm$ 0.4\\
33& 08 25  40  &  -43  10  36  & 15 $\pm$ 4\\
34& 08 25  40  &  -43  31  29    & 18 $\pm$ 8\\
35& 08 25  53  &  -43  24  14    & 15 $\pm$ 5\\
36& 08 26  05 &  -43  22  59.4  & 100 $\pm$ 5 & -1.3 $\pm$ 0.5\\
37& 08 26  11.7 &  -42  39  53.3  & 127 $\pm$ 8 & -0.9 $\pm$ 0.1\\
38& 08 26  30  &  -43  24  43    & 41 $\pm$ 5\\
\hline
\end{tabular}
\begin{list}{}{}
\item{$^a$ The flux density at 1.3 GHz is estimated on the single 128 MHz band 
image (see last paragraph of \S \ref{spind}).}
\end{list}
\end{table}

Here, we refine the calculation of the spectral indices for all five catalogued
sources by two methods: T-T plots and a linear fit to the logarithmic plot 
$S_\nu$ versus $\nu$, where $\nu$ is each of the frequencies in which the 
CABB data were divided, and $S_\nu$ is the flux density of the source. In the 
first case, we used the maps filtered with visibilities between 0.5 and 4 
k$\lambda$ (see 
\S \ref{spind}), and averaged the spectral indices obtained by exchanging both 
frequencies alternately as abscissa or ordinate. To apply the second method, 
flux densities were obtained over the images per individual band (as described 
in the last paragraph of \S \ref{spind}) by fitting a two-dimensional Gaussian 
to each source and integrating the fitted function, allowing for a background 
level. To ensure the robustness of the fits, we made use of two image 
processing packages: {\sc miriad} and {\sc aips}. Both results were in general 
agreement, and the error bars are consistent with the slight differences we 
measure between the results of these two packages.

In Table \ref{cat-cs}, we list all five sources and their spectral indices as 
measured with T-T plots (fourth column) and with the slope of the $S_\nu$ 
versus $\nu$ curve (fifth column). In all cases, the quoted errors correspond 
to the linear fits, regardless of the image sensitivity or other sources 
of uncertainty, like the limits of the integrated regions in the estimation of 
$S_\nu$ or the background level (we adopted the integrated flux density 
provided by the Gaussian fit over the source in each image). The plots and
the corresponding fits are displayed in Fig. \ref{plots-cs}.

\begin{figure}
\centering
\includegraphics[width=.26\textwidth]{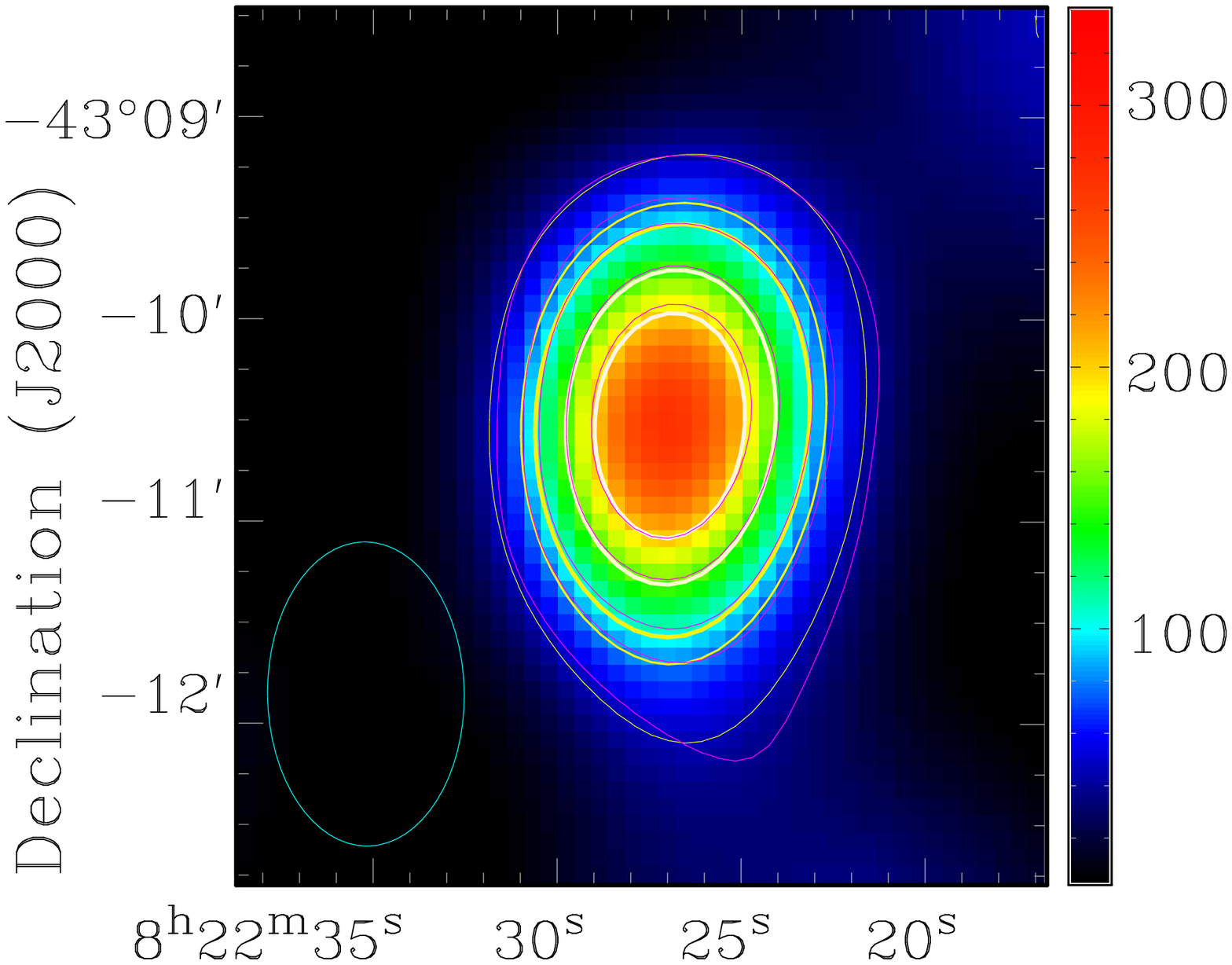}~\hfill 
\includegraphics[width=.22\textwidth]{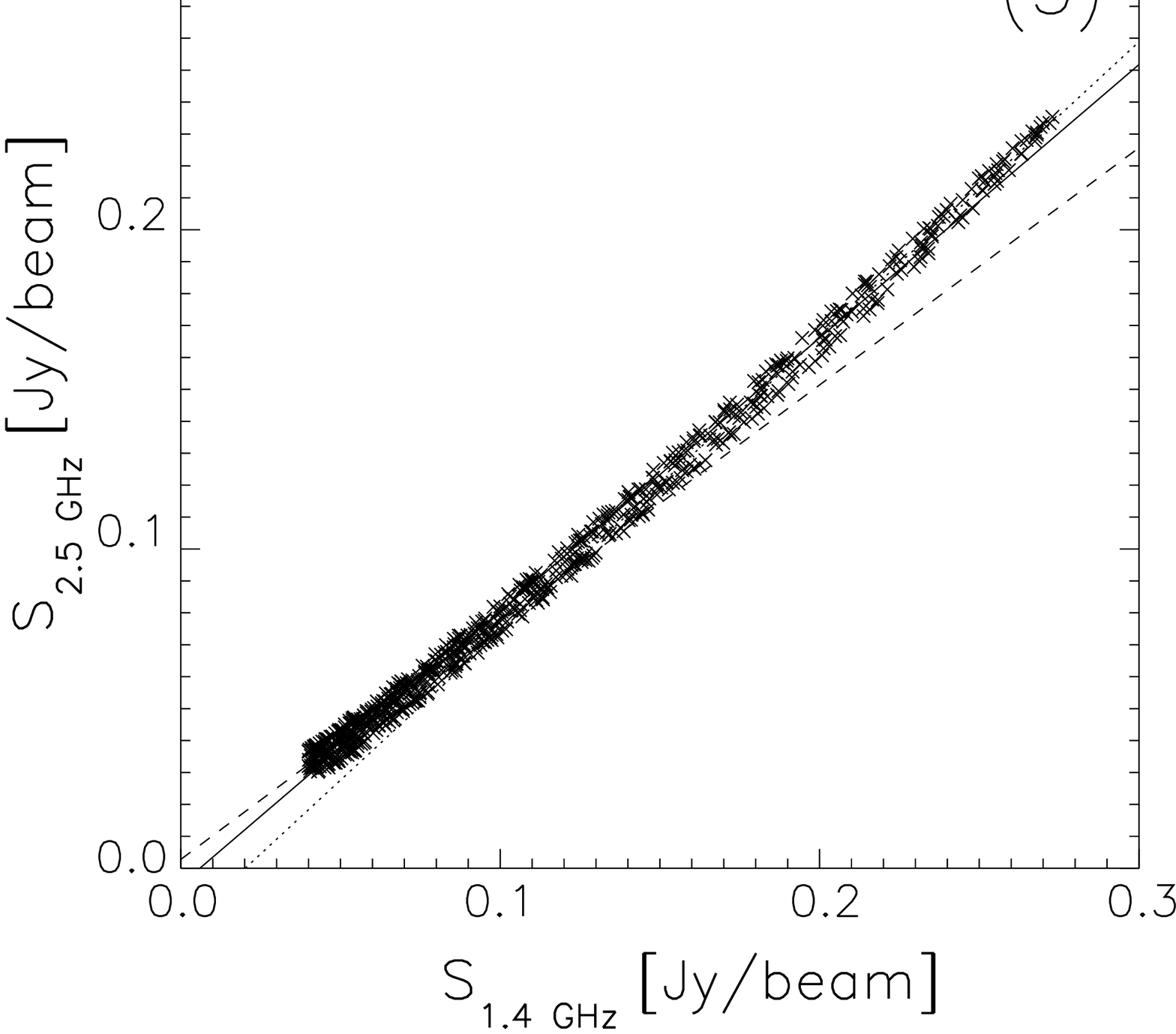}\\
\includegraphics[width=.26\textwidth]{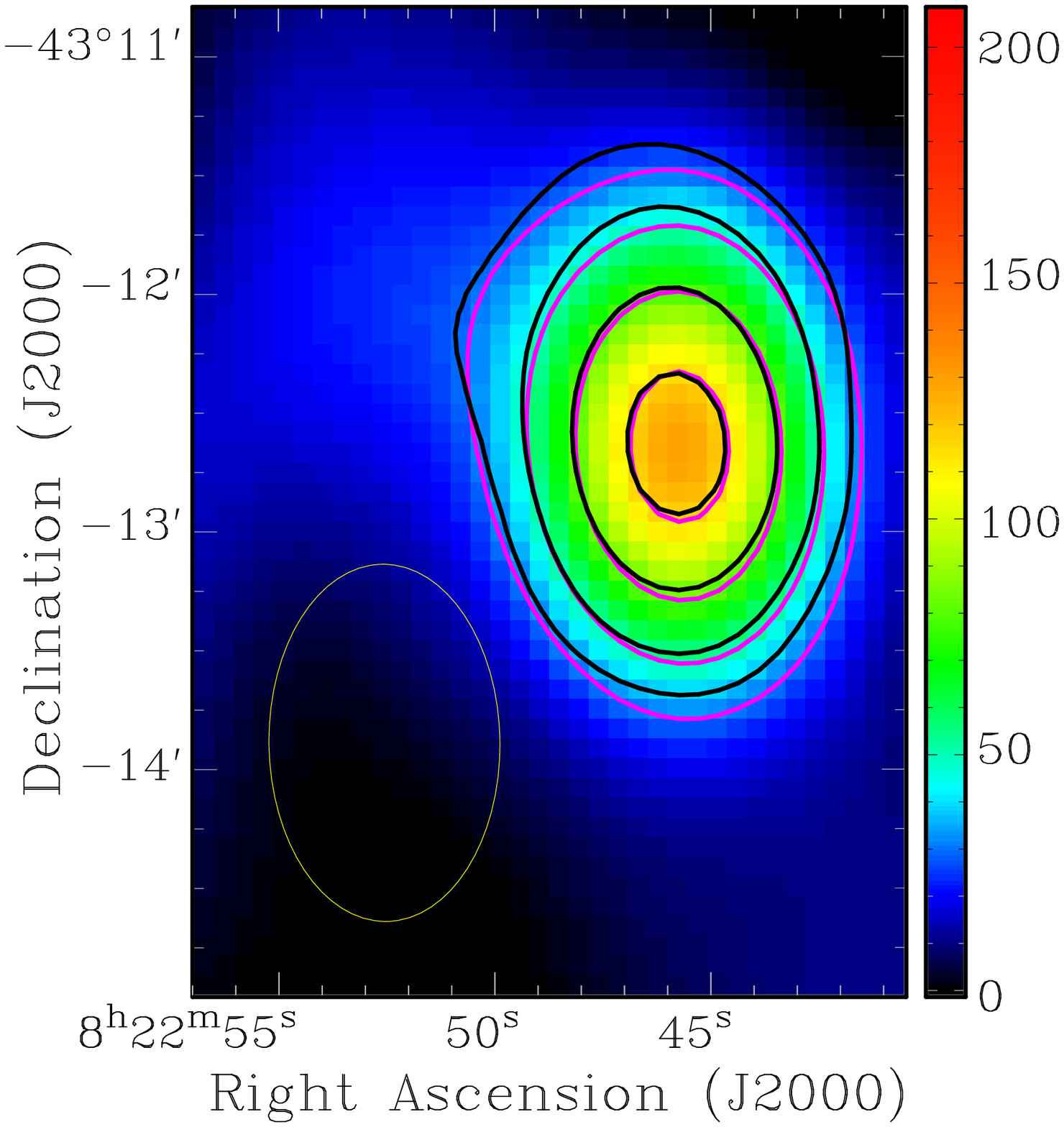}~\hfill  
\includegraphics[width=.22\textwidth]{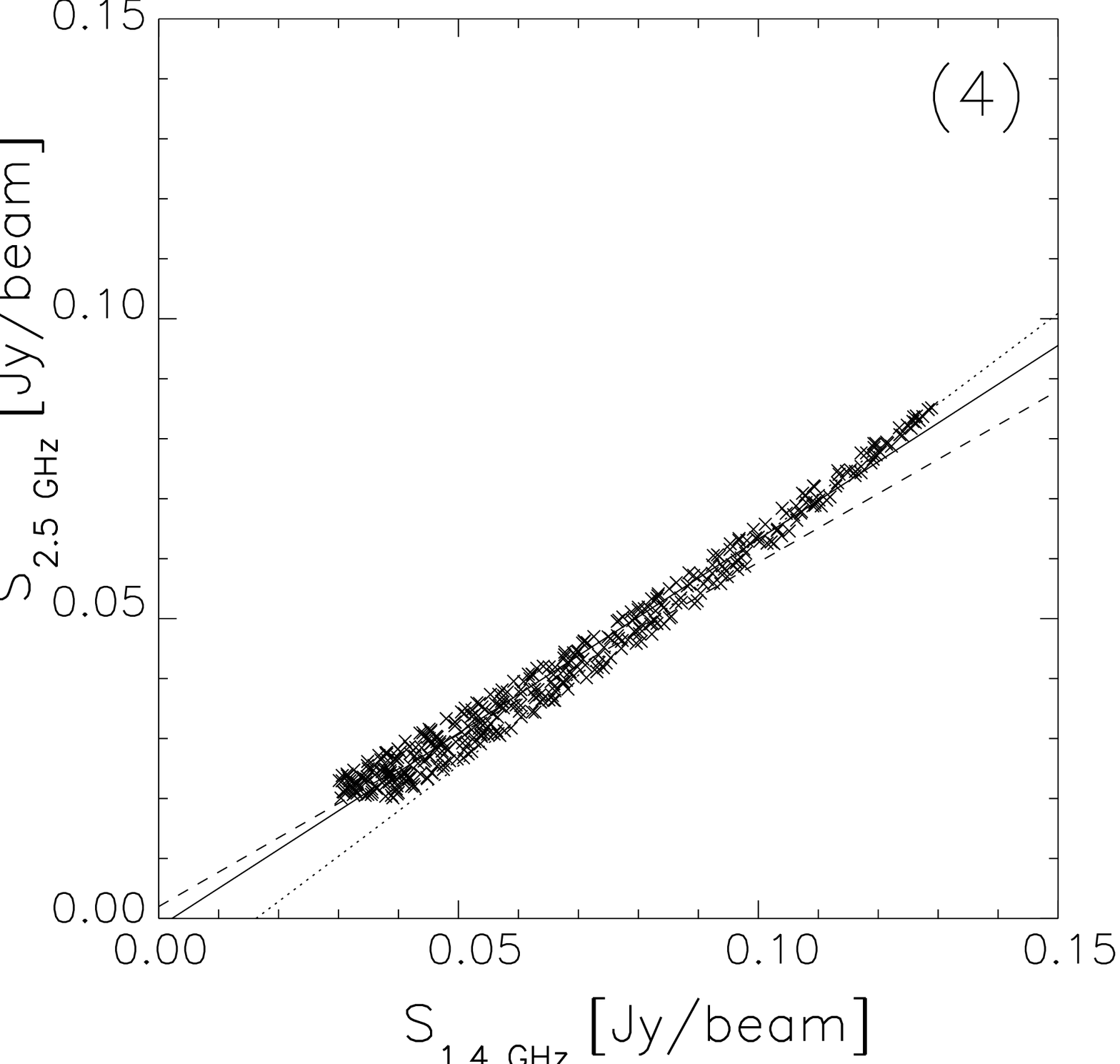}\\
\caption{Top, left-hand panel: compact source 3 at 1.4 GHz. The beam is
plotted at the bottom left corner. Black contours (yellow in the online colour 
version) at 45, 80, 100, 150 and 200 mJy beam$^{-1}$ are overlaid to represent 
the emission at 1.4 GHz, while  grey (magenta) contours correspond to the 
emission levels 35, 60, 80, 120 and 160 mJy beam$^{-1}$ at 2.5 GHz. The flux 
density scale, in units of mJy beam$^{-1}$, is shown at the right. Top, 
right-hand panel: T-T plot for source 3 with the same fit as in Fig. 
\ref{plots-cs} (solid line) and using different flux density ranges (see text). 
Bottom, left-hand panel: compact source 4 at 1.4 GHz. The beam is plotted at 
the bottom left corner. Grey (magenta) contours at 30, 50, 80 and 115 mJy 
beam$^{-1}$ are overlaid to represent the emission at 1.4 GHz, while white 
(black) contours correspond to the emission levels 20, 30, 50 and 75 mJy 
beam$^{-1}$ at 2.5 GHz. The flux density scale, in units of mJy beam$^{-1}$, is 
shown at the right. Bottom, right-hand panel: T-T plot for source 4 with the 
same fit as in Fig. \ref{plots-cs} (solid line) and using different flux 
density ranges (see text). 
}
\label{curvas}
\end{figure}

It is striking that T-T plots show a curvature for sources 3 and 4, with
the highest flux density points departing from the fitted line. To
illustrate this issue, we try different fits to the points and show the
results in Fig.  \ref{curvas}. The left-hand column displays images of sources 
3 (top) and 4 (bottom) at 1.4 GHz constructed with visibilities between 0.5 
and 4 k$\lambda$. The right-hand column shows the corresponding T-T plots, 
where the solid lines coincide with the fits shown in Fig. \ref{plots-cs}, in 
which all flux densities above 45 and 35 mJy beam$^{-1}$ at 1.4 and 2.5 GHz, 
respectively for source 3, and above 30 and 20 mJy beam$^{-1}$ at 1.4 and 2.5 
GHz, respectively, for source 4, were taken into account. Limiting the range up 
to 100 and 80 mJy beam$^{-1}$ at 1.4 and 2.5 GHz, respectively, for source 3, 
and up to 80 and 50 mJy beam$^{-1}$ at 1.4 and 2.5 GHz, respectively, for source
4, the resulting spectral indices are $\alpha = -0.5 \pm 0.3$ for source 3,
and $\alpha = -1.0 \pm 0.4$ for source 4. In Fig. \ref{curvas}, these
fits are represented with a dashed line. If only the brightest core of each 
source is considered (fluxes above 150 and 120 mJy beam$^{-1}$ at 1.4 and 2.5 
GHz, respectively, for source 3, and above 80 and 50 mJy beam$^{-1}$ at 1.4 and 
2.5 GHz, respectively, for source 4), the resulting spectral indices
are $\alpha = -0.14 \pm 0.20$ for source 3, and $\alpha = -0.55 \pm 0.35$ for 
source 4 (dotted lines in Fig. \ref{curvas}). For clarity, the parameters used
for all the fits above and the corresponding spectral indices derived are 
summarized in Table \ref{curv-fits}. The curious behaviour depicted by these 
two sources implies that there is a gradient of spectral indices from the 
centre (flatter) to the edges (steeper).

For completeness, in Table \ref{new-cs} we list all new compact sources 
detected in the field with a flux density above 10 mJy at 1.3 GHz, 
estimated on the single 128 MHz band image (see last paragraph of \S 
\ref{spind}). These sources are numbered following Table \ref{cat-cs}, 
ordered with increasing Right Ascension. In the fourth column, we give the 
integrated flux density at 1.3 GHz. As in the previously catalogued sources, 
a 2D-Gaussian fit was applied to each source, allowing for an offset, and the 
flux density is computed by integrating the fitted distribution. We also 
include a spectral index estimation based on T-T plots obtained as in Table 
\ref{cat-cs} for those sources for which the correlation coefficient of the 
linear fit is better than 0.95.

\section{Discussion}\label{Disc}

\begin{figure}
\centering
\includegraphics[scale=0.5]{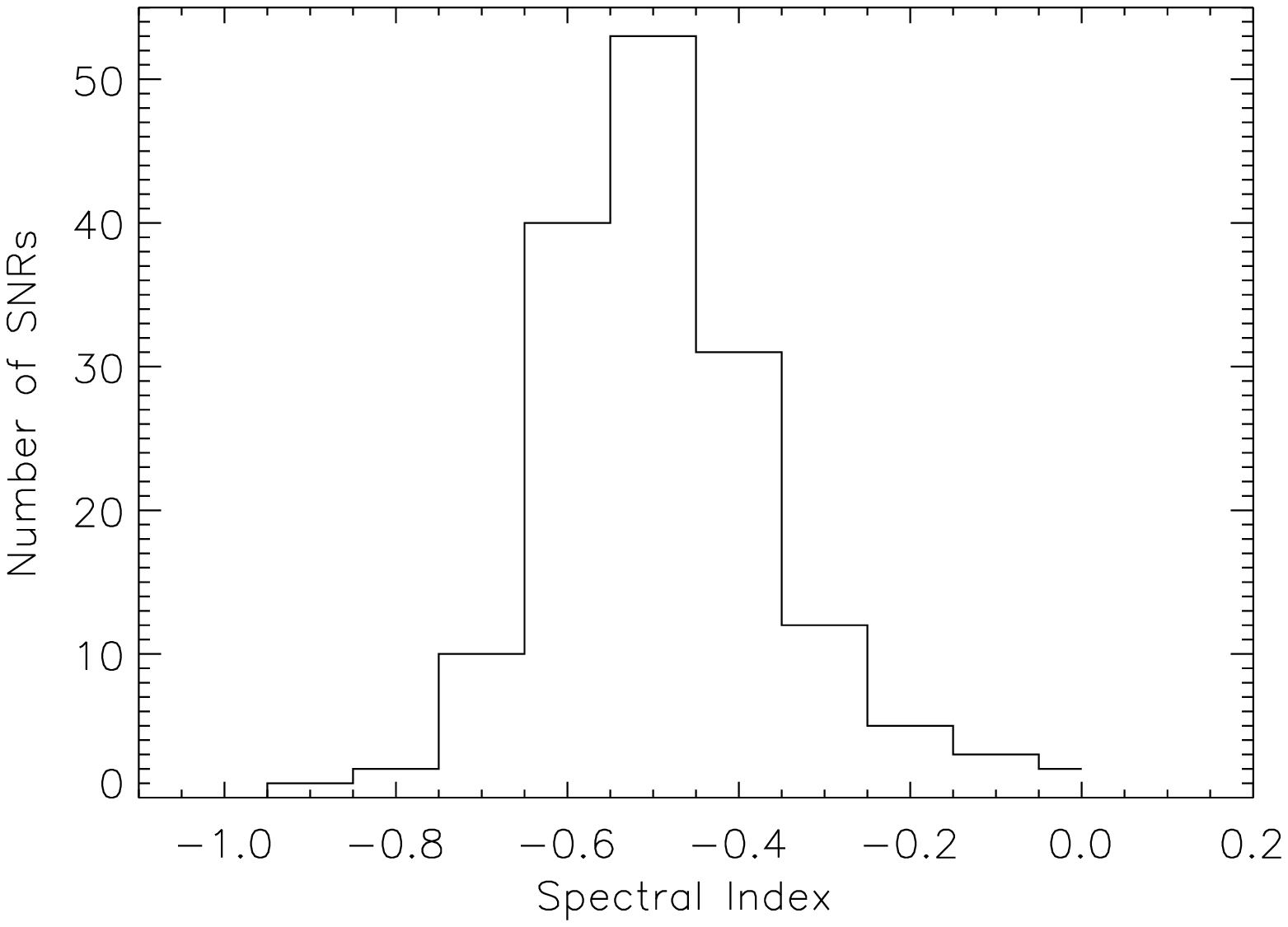}
   \caption{Distribution of the spectral indices of catalogued Galactic SNRs
\citep{BASI}. Only 159 out of the 294 catalogued SNRs were considered, after 
discarding variable or unconfirmed spectral indices.
		}
     \label{histo}
\end{figure}

The energy distribution of an SNR contains information on the mechanisms
involved in particle acceleration processes. Spectra produced by 
shock-accelerated radio-synchrotron emitting electrons typically have {\bf a} 
spectral index $\alpha \simeq -0.6$ \citep[e.g.][]{bkw01}. In the test-particle
limit, adiabatic shocks have a theoretical spectral index $\alpha = -0.5$,
assuming a compression ratio $r=4$ in a monoatomic gas \citep[$\gamma = 5/3$; 
e.g.][]{SReyn01}. In radiative shocks, the test-particle limit is no 
longer valid as accelerated particles influence the shock dynamics, and
can increase the compression ratio and produce flatter spectra. Fig.
\ref{histo} displays the distribution of spectral indices as obtained
from the Catalogue of Galactic Supernova Remnants compiled by \citet{BASI}.
Although the catalogue lists 294 sources, 135 of them have variable spectral
indices, or an undetermined value (quoted value followed by a question mark).
In Fig. \ref{histo}, only the remaining 159 SNRs are considered. Spectral
indices as extreme as 0.0 or $-0.9$ are marginally attained, but most SNRs
have $-0.6 \la \alpha \la -0.4$. Spectral indices can also be altered by
contamination with thermal Bremsstralhung emission, interaction of the shock 
front with a denser medium or high energetic particles injected by an interior 
pulsar.

The most recent determination of the radio spectral index of Puppis A has
been carried out by \citet{gabi+06} who combine a number of flux densities 
corresponding to frequencies between 19 and 8400 MHz, and obtain $\alpha = 
-0.52 \pm 0.03$. Although in agreement within the error limits, this value
is somewhat flatter than the value inferred here from the T-T plot (Fig. 
\ref{TT-Pup}). However, we notice that there appears to be a flattening near 
10-100 MHz \citep[fig. 4 in][]{gabi+06}, where the observational points 
do not follow the fitted single power law model. If the two fluxes at the 
lowest frequencies are excluded, then $\alpha = -0.56 \pm 0.04$, in excellent 
agreement with our result.

Extended SNRs such as Puppis A provide a unique chance to analyse local
variations of the spectral index, which may reflect local changes in the
particle acceleration mechanisms. In general, regions of diffuse emission
have steeper spectra \citep[e.g. Cygnus Loop;][]{uy+04}. As an extreme case,
S147 has a global spectral index of -0.3 but in selected diffuse regions 
it varies between --1.1 and --1.56 \citep{xiao}. The difference between the
spectral indices for bright filaments and diffuse regions in S147 is reflected 
in the two branches where the points gather in the T-T plots. In Puppis A, in 
contrast, all points in the T-T plot can be fitted by a single line (Fig. 
\ref{TT-Pup}), which means that this SNR is well represented by a unique 
global spectral index. Nevertheless, spatial variations are observed, which 
will be analysed in the following section.

\subsection{Spatial spectral index variations}\label{Spvar}

The first study of the spatial distribution of spectral indices in Puppis A
was performed by \citet{gd+91}. They compare two images, at 327 and 1515 MHz,
obtained with the Karl G. Jansky Very Large Array (VLA) and integrate the 
emission in a few boxes over relevant areas. They find that the eastern region 
is considerably steeper than the southern, western, northern, and central 
regions, with $\alpha = -0.67 \pm 0.08$. 

More recently, \citet{gabi+06} improved the analysis by combining new 
observations at 1425 MHz undertaken with the VLA with the old 327 MHz data 
re-processed from scratch. To ensure that all scale structures are sampled,
they integrate single dish observations in their maps. They apply two
techniques to survey the spectral index distribution. The first method
divides directly the logarithm of the ratios between fluxes and frequencies:
\begin{equation}\hskip 3 cm
\alpha = \frac {  {\rm ln} \; {(\frac{S_{\nu_1}}{S_{\nu_2}})}} { {\rm ln} \; 
(\frac{\nu_1}{\nu_2})}.
\end{equation}
The second method, called `tomography', consists of scaling one of 
the images by a test spectral index $\alpha_t$ and subtracting it from 
the other frequency image. The resulting images, $S_{\rm t} = S_{\nu_1} - 
(\frac{\nu_1}{\nu_2})^{\alpha_{\rm t}} S_{\nu_2}$, will be closer to zero as
the test spectral index approaches the real one, and features with
different spectral indices will be highlighted as light (positive) or
dark (negative) as their spectra are steeper or flatter than the assumed 
$\alpha_{\rm t}$. In contrast to \citet{gd+91}, \citet{gabi+06} point out that 
the bright eastern features are somewhat flatter than the rest of the shell.

Both methods are sensitive to offsets in the background emission. In
particular, to include the large-scale emission at 327 MHz, \citet{gabi+06}
use single dish archival data at 408 MHz and scale it with a spectral index 
$\alpha = -0.6$ derived from two images, at 86 MHz \citep{mills60} and at 
408 MHz \citep{AJG71}. Therefore, the subsequent analysis can be influenced to 
some extent by this assumed uniform spectral index. Here, we apply a third 
method which is independent of the background emission level and does
not use archival observations or pre-determined results. The method, described 
in detail in Section \ref{spind}, is called `T-T plots', and the spectral
index distribution is shown in Fig. \ref{alfa-map}. We first call to 
attention that this method does not provide information on the spectral 
index of the interior diffuse emission, since the {\em uv}-filtering performed 
to construct the maps from a same spatial frequency range in the Fourier 
domain, filters out the underlying large-scale emission. Besides, the 
size of the region adopted to compute the fit sets an additional limit 
to the scale to which this method is sensitive, since only spectral indices 
associated  with small-scale features whose brightness vary significantly 
within the region can be measured \citep{AR1993}.

\begin{figure}
\centering
\includegraphics[scale=0.42]{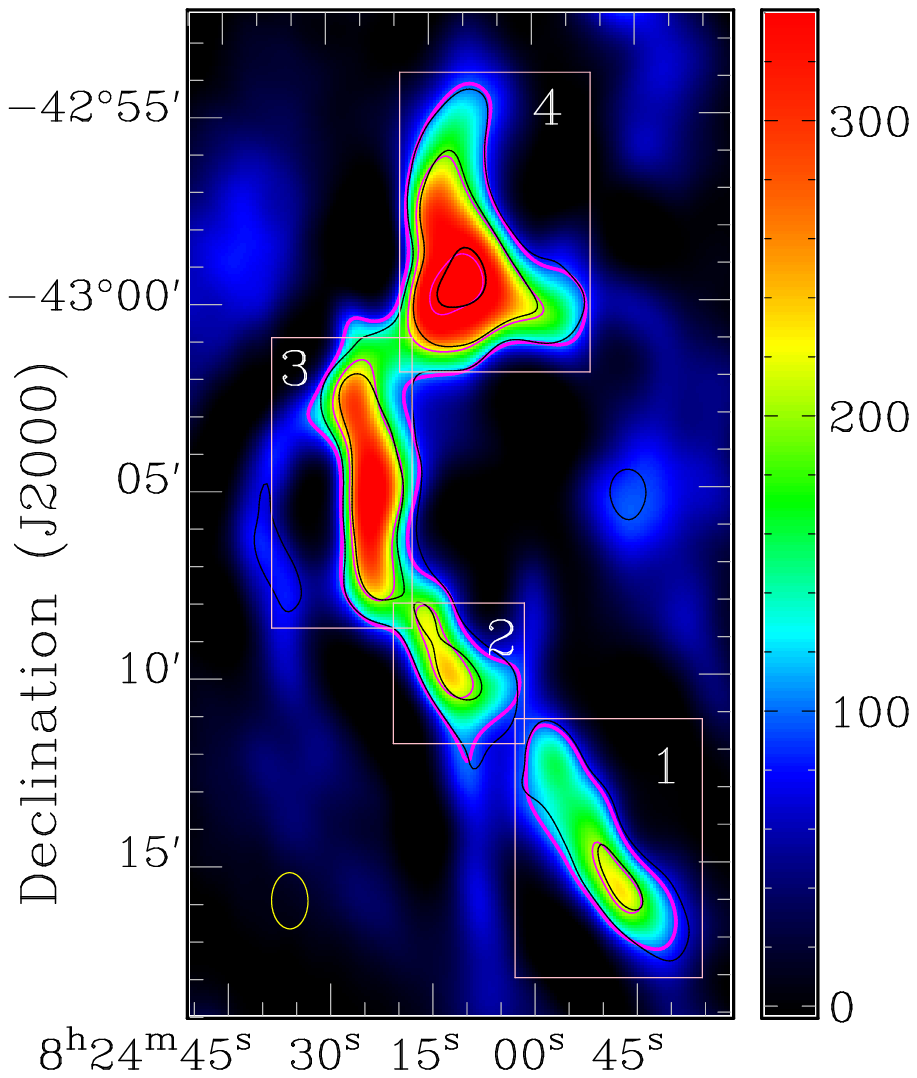}	
\includegraphics[scale=0.32]{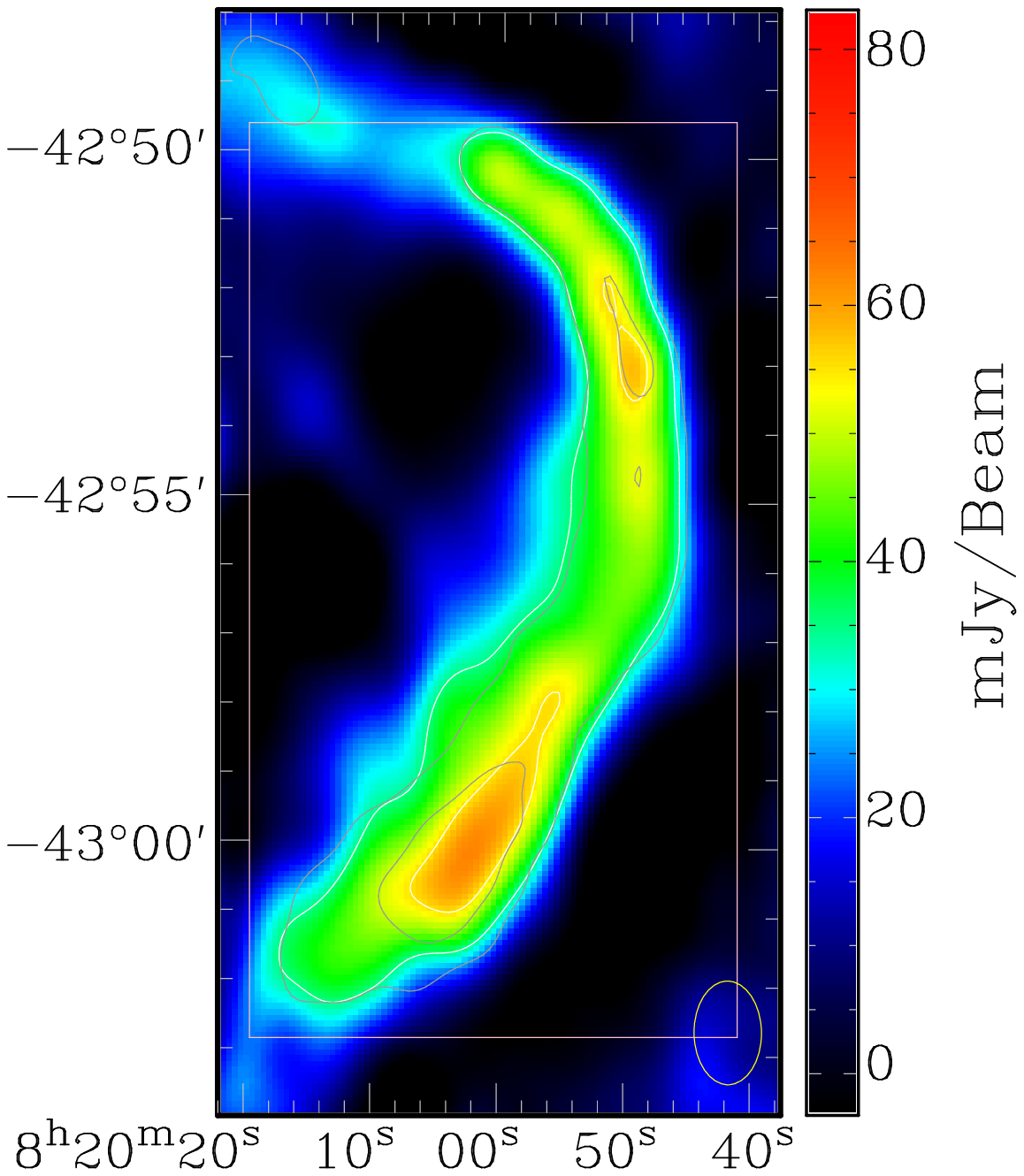}\\	
\vskip 0.30truecm
\includegraphics[scale=0.5]{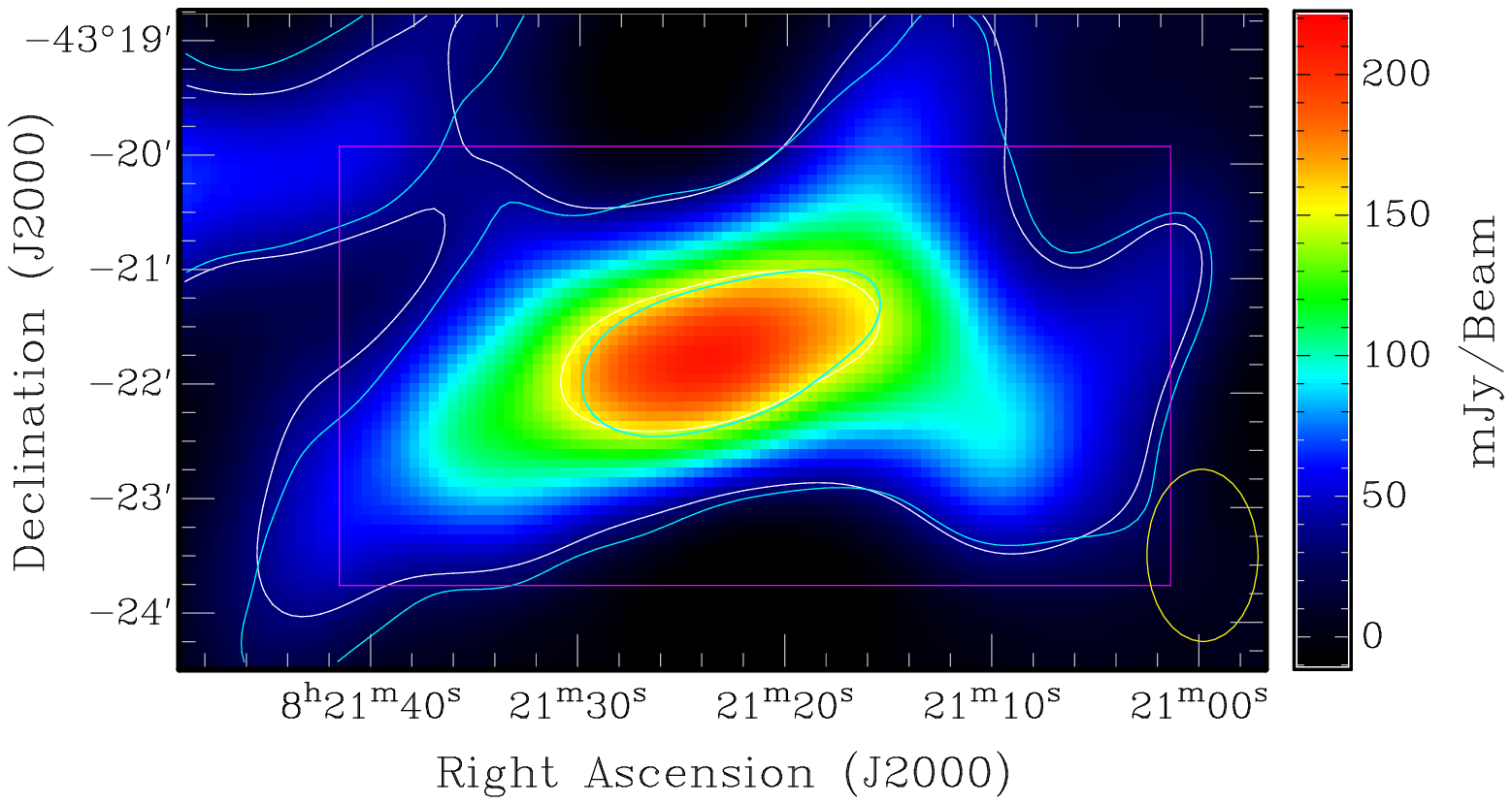}	
   \caption{Radio images of selected regions of Puppis A (top, left-hand panel:
eastern region; top, right-hand panel: western ear-like feature; bottom: 
southern region) at 1.4 GHz using visibilities within the spatial frequency 
range from 0.5 to 4 k$\lambda$. The beam is shown at the bottom. Representative 
contours at 1.4 and 2.5 GHz are overlaid, together with the boxes on which the 
T-T plots were calculated.
		}
     \label{TTp-regs}
\end{figure}

To compare our results with previous studies of spectral index distribution,
we also calculated T-T plots for the regions shown in Fig. \ref{TTp-regs}. For
the southern region (bottom panel), the T-T plot for the area enclosed by
the outermost contour, within the selected box, yields $\alpha = -0.66 \pm
0.20$, steeper than the value reported by \citet{gd+91}. The inner peak has 
a steeper spectrum: $\alpha = -0.8 \pm 0.4$, in agreement with the tomographic 
image in \citet{gabi+06} for $\alpha_{\rm t} = -0.8$, in which most of Puppis A 
has vanished but the southern peak is still visible. The western ear-like 
feature (top, right-hand panel), for which \citet{gd+91} estimated a spectral 
index of --0.56, could not be reliably fitted: we found $\alpha = -0.6$, but 
with a very poor correlation coefficient (lower than 0.75). The low emission at 
this region, close to the rms level, must be a reason for the increase of the 
noise of the T-T plot. Besides, as \citet{gabi+06} noticed, this feature has a 
combination of different structures with variable spectral indices (see also 
Fig. \ref{alfa-map}). 

The eastern region (top, left-hand panel in Fig. \ref{TTp-regs}) has a global 
spectral index of $\alpha = -0.6 \pm 0.1$ when the emission included in the 
outermost contour for the whole map is considered. This value is in agreement 
with \citet{gd+91} within the error margins. If the individual boxes are taken 
separately, the results are: $\alpha_{\rm box\, 1} = -0.60 \pm 0.45$, 
$\alpha_{\rm box\, 2} = -0.64 \pm 0.35$, $\alpha_{\rm box\, 3} = -0.68 \pm 
0.25$ and $\alpha_{\rm box\, 4} = -0.55 \pm 0.12$. Curiously, when the T-T 
plots are limited to the brightest regions, the result is different for boxes 
3 and 4: for the first one, the spectrum steepens ($\alpha = -0.8 \pm 0.4$), 
while for the latter, it gets flatter ($\alpha = -0.50 \pm 0.35$). This 
means that the synchrotron emission and the radio spectral behaviour are not 
necessarily correlated. 

\begin{figure}
\centering
\includegraphics[scale=0.5]{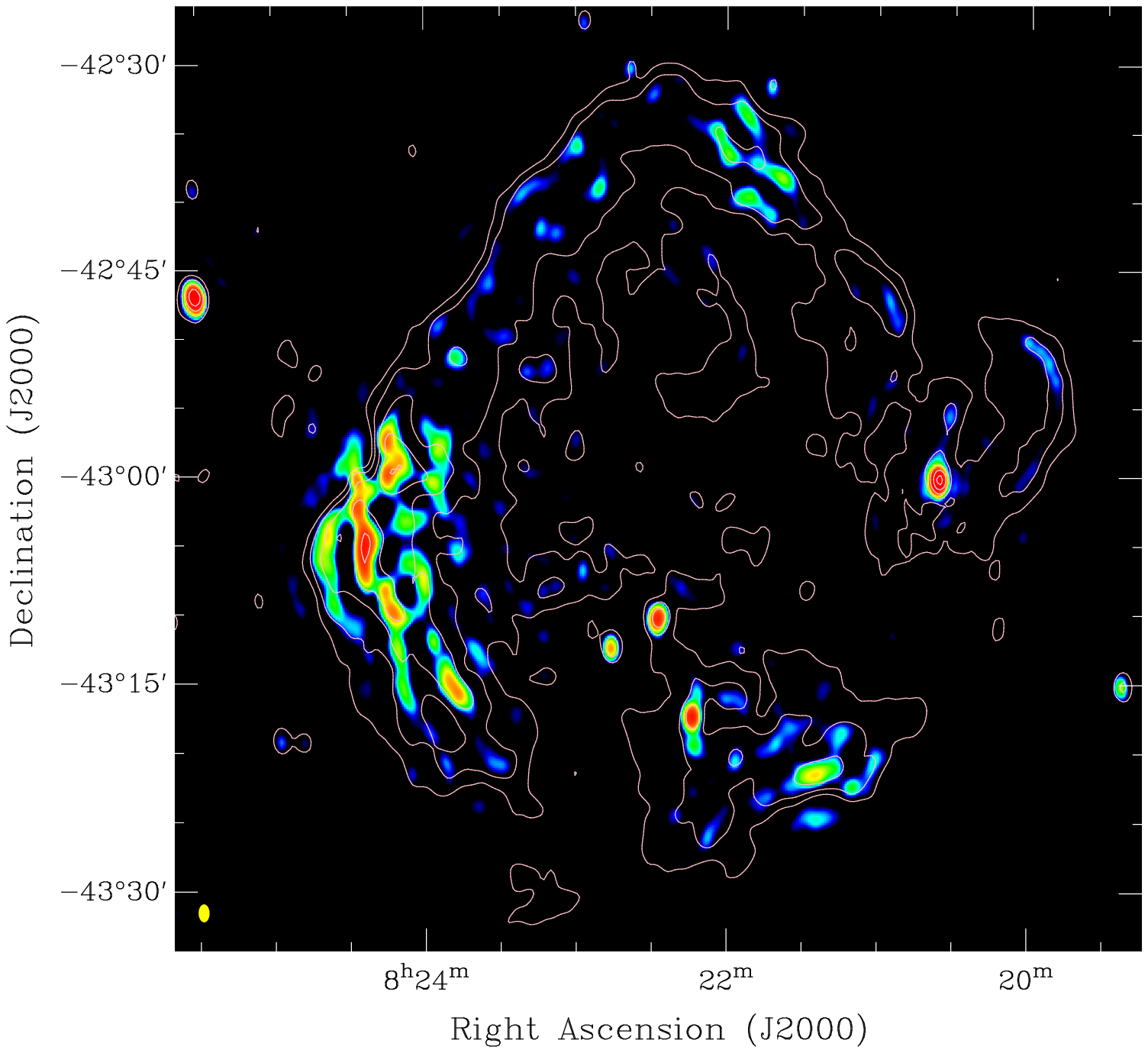}	
   \caption{Radio image of Puppis A at 1.4 GHz using visibilities within the 
spatial frequency range from 1 to 4 k$\lambda$. The beam is shown at the bottom 
left corner. To enhance bright features, the intensity is displayed with a 
logarithmic scale. Contours are as in Fig. \ref{alfa-map}.
		}
     \label{hires}
\end{figure}

In the spectral index map presented in Fig. \ref{alfa-map}, T-T plots were
computed on boxes of $\sim 100 \times 100$ arcsec, which is more than twice
the beam area. In most cases, regions where the spectral indices are more 
extreme (black and white in Fig. \ref{alfa-map}, or blue and red in the online
colour version) are correlated with the larger errors in the spectral index 
determination ($\Delta \alpha \simeq 0.6$). However, the uncertainties of the 
flat spectrum filaments at the south-eastern region are low enough to ensure 
their reliability. Our results do not reproduce the striation found by 
\citet{gabi+06} along the shell, with fringes perpendicular to the Declination 
axis. Instead, as mentioned above, Fig. \ref{alfa-map} shows a pattern of 
narrow strips at the southeastern region parallel to the rim which is not 
correlated with the emission observed in the radio maps (Fig. \ref{ContMos} and 
{\em uv}-filtered images for T-T plots). To explore the nature of these 
filaments, in Fig. \ref{hires} we show an image of Puppis A constructed from a 
more restrictive {\em uv} range, with visibilities from 1 to 4 k$\lambda$, in 
which extended structures are further removed. The eastern region depicts a 
number of bright filaments whose general appearance resembles the pattern 
observed in the spectral index distribution, suggesting that the variations in 
$\alpha$ are connected with structures at this scale. The fact that they do not 
share a detailed correlation, as noticed also by \citet{gabi+06} in Puppis A 
and by \citet{AR1993} in the SNRs G39.2--0.3 and G41.1--0.3, probably means 
that the variations in spectral index are due to aspects of the hydrodynamic 
flow unrelated with synchrotron emissivity. Besides, if the spectrum of the 
electron population is intrinsically bent, different degrees of compression can 
result in spectral index variations \citep{AR1993}.

\begin{figure}
\centering
\includegraphics[scale=0.5]{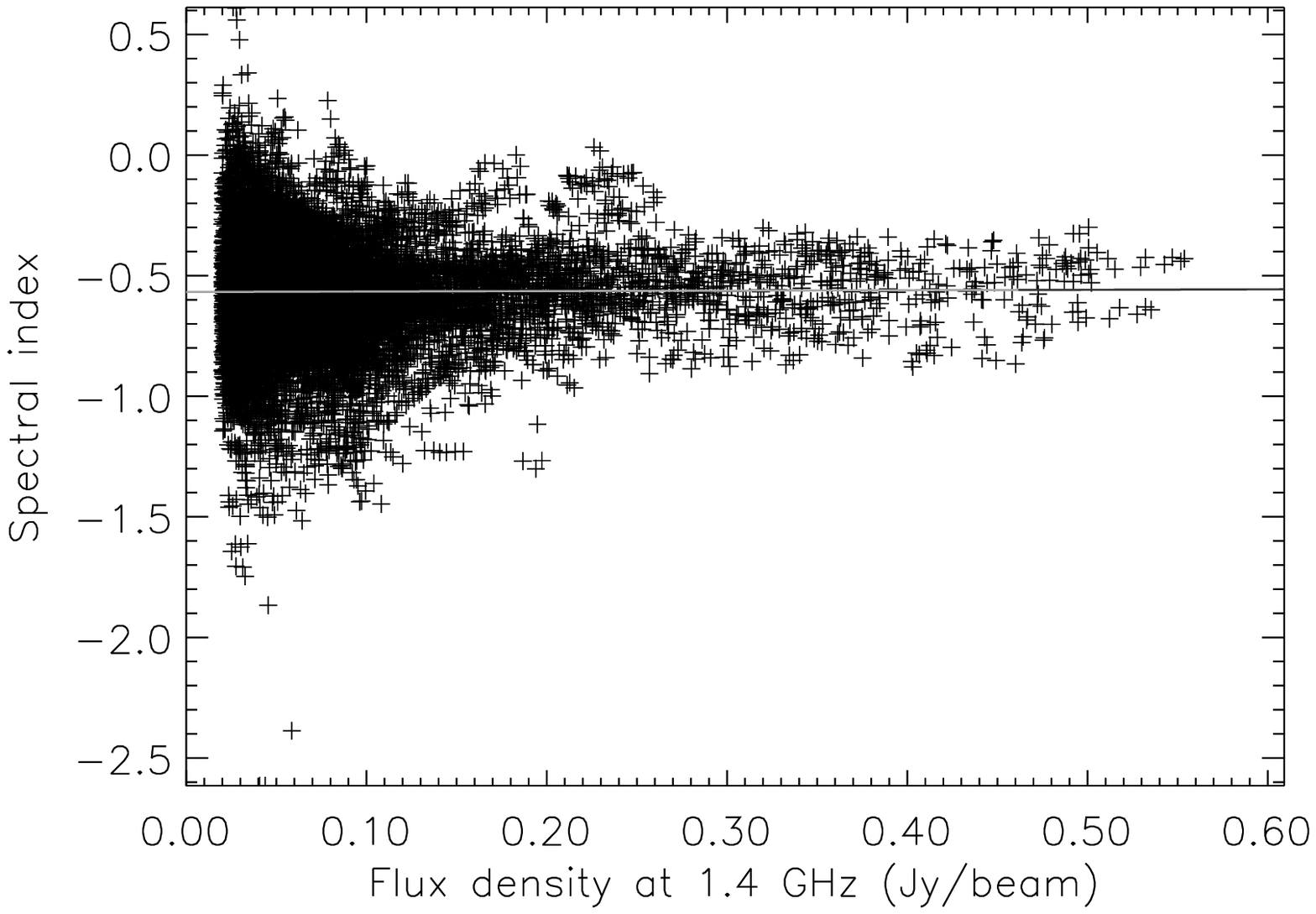}
   \caption{Spectral index versus flux density at 1.4 GHz for Puppis A. Compact
sources have been removed. The solid line represents a linear fit for
flux densities above 20 mJy beam$^{-1}$.
		}
     \label{S-alf}
\end{figure}

\citet{AR1993} mention that in old shell SNRs, the higher emissivity parts are
flatter than the fainter or diffuse parts of the remnants \citep[e.g.][]{fr1986,
uy+04,Ing+14}, while in younger SNRs like Puppis A, the brighter regions 
apparently tend to be associated with steeper indices. To see if there is 
evidence of any such trend, in Fig. \ref{S-alf} we plot the spectral index 
against the flux density of Puppis A at 1.4 GHz after removing the compact 
sources. For flux densities above 20 mJy beam$^{-1}$, the distribution is fit 
by an almost horizontal line, with a slope of 0.018 $\pm$ 0.026. The average 
spectral index is $-0.57 \pm 0.05$, with no variations along the flux density 
axis. This result confirms that there is no obvious correlation between 
emissivity and spectral index.

Flatter spectra are interpreted as sites of active particle acceleration
\citep{AR1996}. As a shock front encounters gas at a higher density and lower
temperature, the Mach number increases and the spectrum flattens \citep{AR1993,
TianL05}. In Fig. \ref{alfa-map}, it is clear that the filament at the 
easternmost part of the shell has a flatter spectrum than the rest of the rim.
This is consistent with the presence of an external cloud interacting with 
Puppis A, as suggested by several atomic and molecular line studies 
\citep{gd+ma88,EMR+95,paron+08}. In other SNRs, shocked clumps are not found
to be associated with flat spectral indices \citep[e.g. HB21;][]{leahy06};
this is probably because any interaction with a denser gas must be followed
by important ionization losses for the spectrum to flatten.

\citet{gabi+06} suggest that the eastern and western `ears' observed in
radio and X-rays could be the termination shocks of two opposite jets ejected 
by the CCO. The spectral index distribution of these features does not offer
any hint to help conclude if this scenario is true, since the eastern ear is 
clearly flatter than the western one. A possible explanation for the difference
in $\alpha$ if both features share the same origin, would be that the eastern 
shock has encountered a denser ISM and the flatter spectral index is due to 
this interaction.

\subsection{Compact sources}\label{Comp-src}

Previous high-resolution radio surveys of Puppis A \citep{MGD83,gd+91,gabi+06} 
have reported the presence of four compact sources superimposed on the SNR 
face\footnote{A Molonglo contour map at 408 MHz \citep{AJG71} shows all four 
sources, but only the brightest one (source 1) is identified as an unresolved 
source with a spectrum steeper than the rest of Puppis A.}, listed as sources 1 
to 4 in Table \ref{cat-cs}. Source 3 is the only one with an X-ray counterpart 
\citep{Xgd+13}. \citet{MGD83} estimate spectral indices for these four sources 
comparing their observations at 21 cm with the Molongo map at 408 MHz
\citep{AJG71} and find $\alpha \simeq -1.0$ for all of them. \citet{gd+91}
compare flux densities at 327 and 1515 MHz and find spectra steeper than 
$-1.0$ for all but source 3, for which $\alpha = -0.8$. Based on the tomographic
image for a test spectral index of $-0.8$, \citet{gabi+06} conclude that all
four sources are steeper than $-0.8$ and, therefore, extragalactic. Our results 
do not confirm these results except for source 2, which is clearly extragalactic, 
together with source 5 which lacks previous estimations of the spectral index. 
However, the results are more in line if we note that source 3 does not appear 
anymore in the tomographic image of \citet{gabi+06} with $\alpha_{\rm t} = -0.8$
(hence the spectral index is limited between $-0.8 < \alpha < -0.6$), and source
4 is marginally observed, which means that $\alpha \simeq -0.8$ but not steeper.

From Table \ref{cat-cs}, spectral indices measured from the curve $S(\nu)$ 
versus $\nu$ are systematically steeper than those derived using T-T plots. For 
sources 1 and 4, the steeper indices would not be in conflict with previous 
results. However, spectral indices obtained with the T-T plot method are more 
reliable, since they are free of the following problems involved with flux 
estimates. In the first place, splitting the {\em uv} data into 128 MHz bands 
has the disadvantage that the reduced number of visibilities per band increases 
the noise level of the images and makes the removal of sidelobes more 
challenging. And secondly, if the 2D Gaussian or offset fits are poor, the 
integrated flux density may not be accurate. The departure of several measured 
points from the fitted lines in the right-hand column of Fig. \ref{plots-cs} 
can be ascribed to these two problems. In the case of sources 3 and 4, we have
discussed in \S \ref{spind-cs} that $\alpha$ varies from the core to the edges.
When the core is excluded, the spectral indices derived from T-T plots and 
by fitting a line to $S(\nu)$ versus $\nu$ are coincident (compare Tables 
\ref{cat-cs} and \ref{curv-fits}). This means that the contribution from the
cores is negligible when computing the total fluxes of these sources.

It is possible that the difference between the values of $\alpha$ measured 
here and in earlier works can be explained because sources 1-4 are superimposed 
on the diffuse component of Puppis A, and previous flux density determinations 
are probably contaminated with this background emission. It is also likely that 
one or more of these sources are variable. As an example, we notice that there 
is a difference of about a factor of 2 between the flux densities of sources 1 
and 2 at 21 cm \citep[500 and 230 mJy, respectively;][]{MGD83} and at 1515 MHz 
\citep[230 and 160 mJy, respectively;][]{gd+91}, whereas according to the 
quoted spectral indices the difference should be less than 10 per cent. In that 
sense, the present work exhibits the unique advantage of being based on 
simultaneous observations.

\begin{figure}
\centering
\includegraphics[width=.24\textwidth]{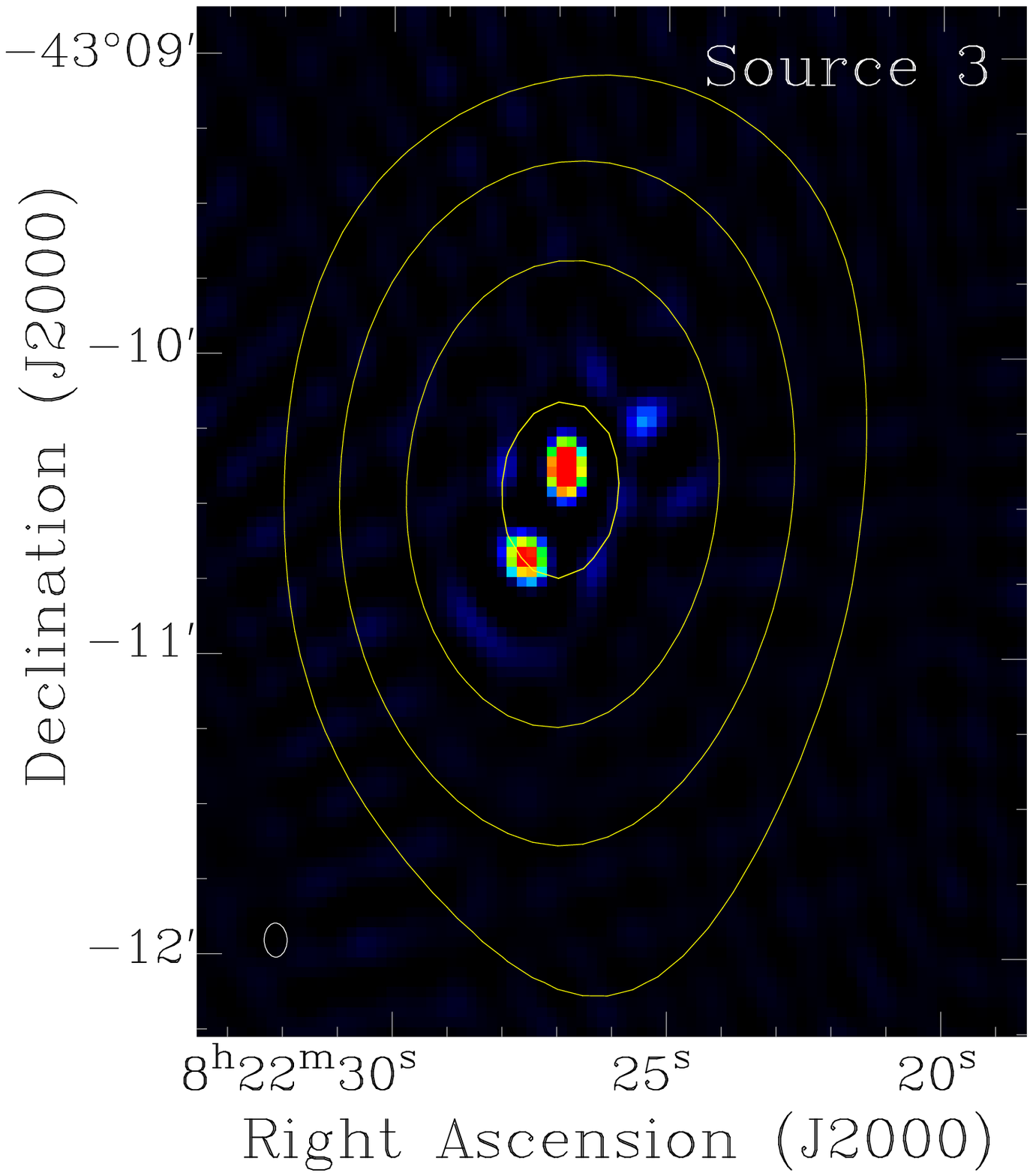}~\hfill
\includegraphics[width=.24\textwidth]{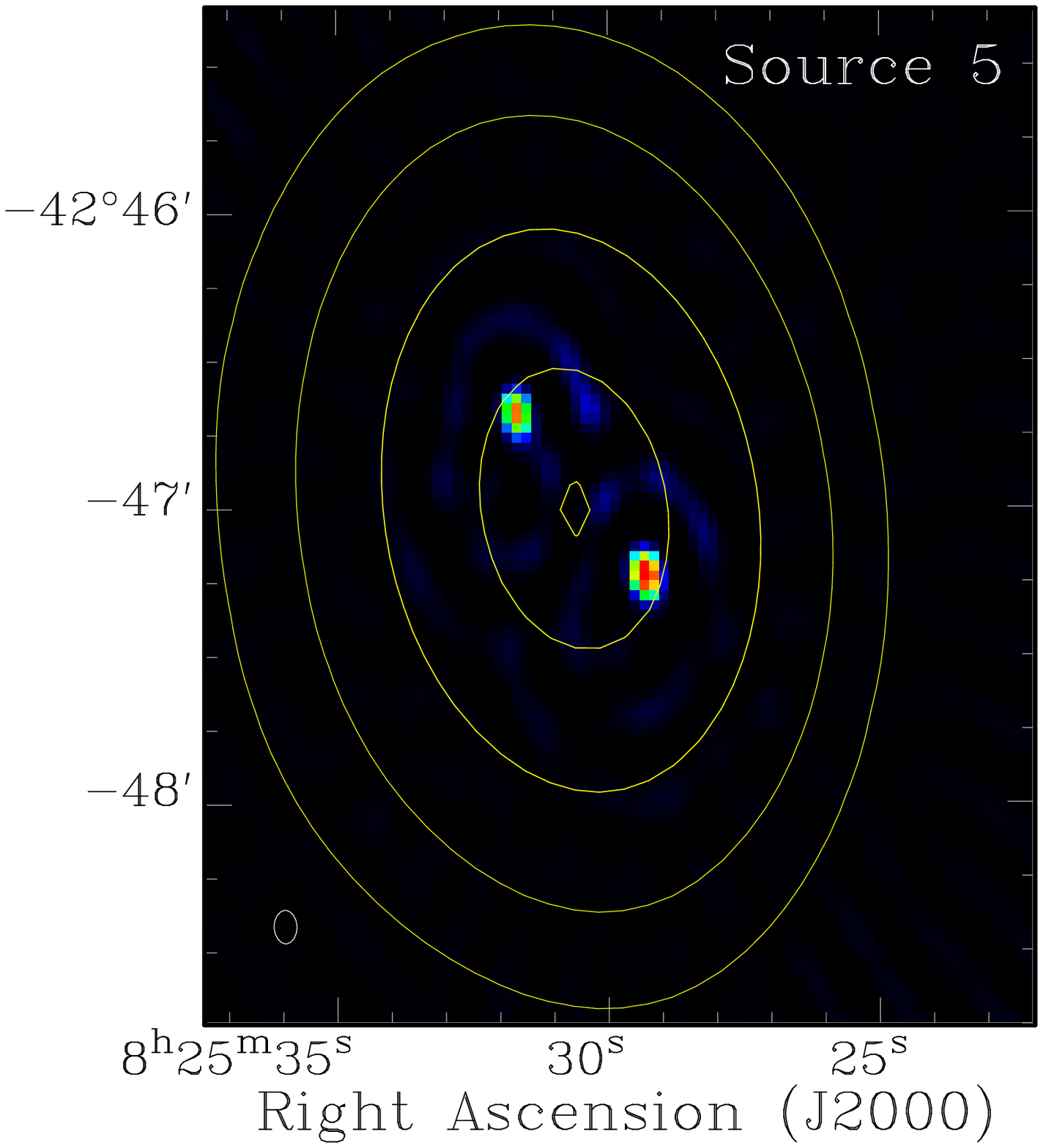}\\
\caption{Left: very high resolution (HPBW = $6.8 \times 4.6$ arcsec) image of 
source 3 at 1.4 GHz. The beam is indicated in the bottom left corner. Contours 
of an image at 1.4 GHz filtered in {\em uv} between 0.2 and 4 k$\lambda$ (\S 
\ref{Obs}) are overlaid, at 40, 80, 150 and 250 mJy beam$^{-1}$. Right: same as 
left for source 5, where the plotted contours are 30, 80, 200, 400 and 500 mJy 
beam$^{-1}$.}
\label{ant6}
\end{figure}

Finally, we have made use of the visibilities associated with the 6th antenna
of the ATCA, which vary in the range between 8 and 24 k$\lambda$, to construct
very high resolution images (HPBW=$6.8 \times 4.6$ arcsec) of the brightest 
sources. These images reveal that sources 3 and 5 are double, as shown in
Fig. \ref{ant6}, where the former could be even triple (a weaker feature 
probably real appears at the NW of the two brighter components).  Such
double-lobed structures are typical of extragalactic sources. The multiple
nature of source 3 could explain the different spectral indices from the edge
to the centre. However, source 4 does not show any structure on small scales,
hence the reason of its spectral behaviour remains a mystery. Although 
previously identified as an extragalactic source, it is more likely a Galactic 
source that may even be associated with Puppis A.

Among the new compact sources catalogued in Table \ref{new-cs}, double sources 
are also: 10, 21, 22, 25, 26 and 38, while source 36 is probably triple. 
Unfortunately, the {\em uv} sampling obtained with baselines that include only 
the 6th antenna is too scarce to perform a deeper investigation into the 
properties of these sources. New observations with the ATCA in its 6 km 
configurations are desirable to uncover the nature of these multiple sources.

\subsection{Ejecta `bullets' in Puppis A?}

There is plenty of evidence that core-collapse SN eject `bullets' of stellar 
material \citep[e.g.][]{WCh2002}. These clumps evolve from an initial bow-shock 
phase to an instability stage, where Rayleigh--Taylor {\bf (R-T)} fingers are 
developed along the forward face \citep{And+94}. The high-velocity optical 
O-rich knots detected in Cassiopeia A \citep{vdBergh71} and G292.0+1.8 
\citep{ghw2005} are SN ejected clumps in the early bow-shock phase. At more 
evolved stages, R-T fingers stretch and amplify the local magnetic field, and 
the clumps become bright radio-synchrotron emitters. Compact radio knots are 
observed in the young SNR Cassiopeia A \citep{And+94} and in the older Vela SNR 
\citep{Duncan+96}. In Cas A, radio knots have typical spectral indices around 
$-0.75$ independently of the knot velocity, and at the outer boundaries of the 
shell, spectra are even steeper \citep[$\alpha \simeq -0.85$;][]{And+94}. In 
Vela, several radio knots are the counterparts of high-velocity X-ray features 
\citep{Strom+95,Duncan+96}. 

\citet{gabi+06} observed the presence of a compact feature (size $\sim 45$ 
arcsec) to the east, close to an intense X-ray region known as the `bright 
eastern knot' \citep[BEK;][]{Petre+82,BEK+05}. They conclude that this source 
is extragalactic based on a spectrum steeper than $-0.8$ inferred from the 
tomographic images. In fact, this source is hardly observed in their tomographic
image traced with $\alpha_{\rm t} = -0.8$, which means that the spectral index 
is limited between $-0.8 < \alpha < -0.6$. We analysed our images filtered with 
visibilities between 1 and 4 k$\lambda$ (Fig. \ref{hires}) and could locate this
source at RA(2000)=8$^{\mathrm h}$24$^{\mathrm m}$13\hbox{$.\!\!{}^{\rm s}$}9, 
Dec.(2000)=$-42^\circ$59$^{\prime}$55$^{\prime\prime}$. The source is resolved, 
and its size is about $86 \times 42$ arcsec after deconvolving the fitted 
Gaussian by the beam. A T-T plot yields $\alpha = -0.8 \pm 1.1$. Thus, not only 
is the proposed extragalactic nature of this source not unequivocal, but the 
spectral index is compatible with a radio knot, like those found in Cas A.

\begin{table}
\caption{Radio knot candidates$^a$} 
\label{knots}
\centering
\begin{tabular}{ccr}
\hline
\hline
 RA (J2000) & Dec. (J2000) & ~Spectral\\
 ~~h~~ m ~~ s & ~~$^\circ$ ~~~$^\prime$ ~~~$^{\prime\prime}$ & index \\
\hline

 08 21  38.3 &  -42  38  22  & -0.8 $\pm$ 0.5\\
 08 21  51.4 &  -42  39  51 &  -0.7 $\pm$ 0.7 \\
 08 22  12.5  &  -43  19  36    & -0.6 $\pm$ 1.1\\
 08 22  50.4 &  -42  39  09  &  -0.8 $\pm$ 0.9 \\
 08 23  18.7  &  -42  52  34    &  -0.8 $\pm$ 1.1 \\
 08 24  12.4  &  -43  09  58    &  -0.6 $\pm$ 1.0\\
 08 24  13.9  &  -42  59  55    &  -0.8 $\pm$ 1.1\\
\hline
 08 21  42.2 &  -42  41  09  & -1.3 $\pm$ 0.6\\
 08 22  59.4 &  -42  36  06  &  -1.1 $\pm$ 0.3 \\
 08 23  07.7 &  -42  42  24  &  -1.3 $\pm$ 0.5\\
 08 23  14.1  & -42  41  56  &   -1.2 $\pm$ 1.2 \\
 08 23  47.3 &  -42  51  33  &  -1.2 $\pm$ 1.6\\
\hline
 08 21  56.0  &  -43  20  51  &  -0.2 $\pm$ 0.7\\
 08 23  56.37  &  -43  12  09    &  -0.35 $\pm$ 0.60\\
\hline
\end{tabular}
\begin{list}{}{}
 \item  $^a$ Radio knot candidates are listed in the upper band. For 
completeness, sources probably extragalactic found in the search are 
included in the central band, and flat spectrum sources, in the lower band.
\end{list}
\end{table}

Remarkably, this compact radio feature is located next to the BEK. 
\citet{BEK+05} interpreted the X-ray emission from the BEK as coming from a 
shocked interstellar cloud based on the `voided sphere' morphology similar to 
that produced in experimental simulations of interaction between shock waves and
dense clouds. However, a molecular study failed to detect any gas concentration
associated with the BEK \citep{paron+08}. It should not be discarded that both 
the BEK and the radio source reported by \citet{gabi+06} are evolved ejecta 
bullets like those detected in X-rays and radio wavelengths in other SNRs. 
Moreover, the coexistence of ejecta clumps in their earlier stages (as optical 
O-rich knots) and in their more evolved phase (as radio knots) in Cas A 
suggests that radio knots can also be found in Puppis A, since both remnants 
share the characteristic system of O-rich high-velocity clumps ejected in the 
SN event. 

We have searched for radio knots along the shell and found six more candidates, 
based on their spectral indices, which are displayed in Table \ref{knots}. We 
also list five compact sources with spectra steeper than -1, hence probably of 
extragalactic origin (although the error margins make their classification not 
conclusive), and two compact sources with a flat spectrum. Higher resolution 
and sensitivity images of these radio knot candidates will help to measure 
changes in intensity and position, and to unveil if they are individual clumps 
or are composed by a cluster of knots. Proper motion studies are crucial to 
conclude if these features are high-velocity ejecta.

\section{Conclusions}

Using the ATCA, we have imaged the prominent SNR Puppis A in continuum emission 
between 1 and 3 GHz. The observations included full 12 hr synthesis imaging in 
two arrays -- EW\,352 and 750\,A. Our images are composed of 24 mosaic pointings
and cover an area approximately $2 \times 2$ deg$^2$. The resultant data are of 
high sensitivity (1.5\,mJy beam$^{-1}$) and high resolution ($\sim 80 \times 
50$ arcsec).

We have used T-T plots to calculate the spectral index across Puppis A 
and derive a value of $-0.563\pm0.013$. Although local variations in the 
spectral index are noted, this global spectral index appears to well 
represent most of the SNR. Contrary to previous suggestions, we find no 
correlation between spectral indices and radio brightness. We identify 
filamentary structure, particularly in the eastern edge of the SNR, which 
resembles the pattern followed by the spectral index variations. This pattern 
was not detected in previous studies. We suggest that the outermost filament 
may be the result of interaction of the SNR with an external cloud. There 
is no evidence, based on the spectral index, of the eastern and western 
`ears' being termination shocks of two opposite jets from the CCO. We also 
found a number of compact features that could be radio knots resulting from the 
evolution of clumpy SN ejecta. A dynamical study should clarify this hypothesis.

We investigate the nature of a number of compact sources that are found 
projected both within and outside of the SNR. Some of these compact sources do 
show strongly negative spectral indices, which are indicative of extragalactic 
sources. It is not clear that sources 1, 3, and 4 are extragalactic, as claimed 
in previous works. With the highest spatial resolution available to us, we are 
able to identify double-lobed structures in sources 3 and 5, confirming the 
extragalactic nature of the latter and suggesting the same for the former in 
spite of its flat spectral index. However, source 4, which was previously 
identified as extragalactic, shows a flatter spectral index with a radial 
gradient (like source 3) and does not show any structure on small scales. We 
therefore expect that source 4 is more likely a Galactic source that may well 
be associated with the SNR.

\section*{Acknowledgements}

We thank the referee (Luke Bozzetto) for helpful and constructive 
criticism, which has improved the quality of this paper.
We appreciate useful discussions with Laura Richter, Timothy Shimwell and 
Mark Wieringa for solving problems during data reduction. This research was 
partially funded by CONICET grants PIP 114-200801-00428 and 112-201207-00226.
The Australia Telescope Compact Array is part of the Australia Telescope 
National Facility which is funded by the Commonwealth of Australia for 
operation as a National Facility managed by CSIRO. EMR is member of the 
Carrera del Investigador Cient\'\i fico of CONICET, Argentina.

\bibliographystyle{mn2e}
\bibliography{apj-jour,puppis}{}

\bsp

\label{lastpage}

\end{document}